%% file: 1-main_ArXiV.tex
\documentclass[11pt,twocolumn]{article}
\usepackage[margin=1in,columnsep=0.35in]{geometry}
\usepackage[utf8]{inputenc}
\usepackage[T1]{fontenc}
\usepackage{microtype}
\usepackage{amsmath}
\usepackage{amsthm}

\usepackage{graphicx}
\usepackage{url}
\usepackage{soul}
\usepackage{xcolor}
\usepackage{colortbl}   
\usepackage{array}
\usepackage[version=4]{mhchem}
\usepackage{siunitx}
\usepackage{subcaption}
\usepackage{booktabs,tabularx,array} 
\usepackage{multirow}
\usepackage{threeparttable} 
\usepackage{comment}
\usepackage{wasysym}
\usepackage[authoryear,round]{natbib}
\usepackage{authblk}
\usepackage{fancyhdr}
\fancypagestyle{firstpage}{
    \fancyhf{}
    
    \fancyfoot[C]{\textit{In Preparation}}
}
\usepackage[colorlinks=true,citecolor=blue,linkcolor=blue,urlcolor=blue]{hyperref} 

\usepackage[switch]{lineno}   

\definecolor{xkcdvioletblue}{HTML}{1D5DEC}
\definecolor{xkcdmossgreen}{HTML}{3D7A3D}
\definecolor{lightgrey}{RGB}{200,200,200}

\makeatletter
\newsavebox{\abstractbox}
\renewenvironment{abstract}
  {\global\setbox\abstractbox\vbox\bgroup
   \hsize=0.9\textwidth \parindent=0pt \small
   \noindent\textbf{Abstract.}\ \ignorespaces}
  {\egroup}
\renewcommand\@maketitle{%
  \newpage\null\vskip 0.5em
  \begin{center}%
    {\large\bfseries \@title\par}%
    \vskip 0.8em
    {\@author\par}%
    \vskip 1em
    \usebox{\abstractbox}%
  \end{center}%
  \par\vskip 0.5em}
\makeatother
\let\cite\citep   
\title{On the Two-Dimensional Structure and Asymmetries of Ionic Liquid Electrospray Plumes}

\author[1]{Zach Ulibarri}
\author[1]{Giuliana Hofheins}
\author[1]{Sophia Gessman}
\author[1]{Elaine Petro}
\affil[1]{ASTRA Lab, Sibley School of Mechanical \& Aerospace Engineering, Cornell University, Ithaca, NY 14850, USA}

\date{}

\begin{document}

\begin{abstract}
        Vacuum electrosprays play important roles in electrospray propulsion, a type of spacecraft electric propulsion wherein charged ions or droplets are accelerated at high velocity to provide spacecraft thrust, and they also have important implications for surface etching and mass spectrometry. Here we present the first fully two-dimensional time-of-flight (TOF) mass spectrometry survey of a vacuum electrospray plume, generated by a tungsten needle externally-wetted with the ionic liquid 1-Ethyl-3-methylimidazolium tetrafluoroborate (EMI-BF$_4$). We find that the plume exhibits clear two-dimensional compositional variation, structure, and asymmetry, with heavy particles and energetic neutrals being more prevalent in the center and a ring-shaped distribution for the monomers (the lightest molecular ions) peaking at 10$^\circ$ off beam-axis with a relative minima in the center. Using a new spatial baseline characterization method, we find a population of energetic neutrals firing 3.4$^\circ$ off beam-axis. In particular, we find that by comparing different parts of the plume, the estimated propulsive efficiency from any one sampled point may vary by as much as a factor of 6. We also find that high mass droplets, which are often assumed as absent in many studies of externally-wetted needles, may carry away significant propellant mass at lower effective velocity and reveal a cone-jet mode of operation at 470 nA output current. We thus find that whole-plume compositional surveys may more accurately assess plume composition and propulsive efficiency, and a significant portion of the `missing mass' in electrospray propulsion sources presumed to be operating in the pure ion regime can be potentially explained by limited sampling of the spatially non-uniform ion plume.

    \end{abstract}

\maketitle
\thispagestyle{firstpage}

	\section{Introduction}
  
    Electrospray thrusters are an emerging electric propulsion (EP) technology that operate by electrostatically firing ions from vacuum stable liquids at exceptionally high velocity \citep{gamero2002electric,hruby2001micro,romero2005ionic}. Room temperature ionic liquids (RTILs) are cation-anion paired salts in liquid form at room temperature, and they are very often stable in vacuum \citep{rogers2007ionic}. The unique properties of these vacuum-stable RTILs makes them attractive propellants for electrospray propulsion, whereby molecular ions or charged droplets are accelerated to high velocity to create spacecraft thrust. Such electrospray propulsion systems have the potential to be among the most efficient thrust sources available \citep{krejci2018space}. They are also capable of providing a level of precise control that is difficult or impossible to achieve with other EP sources, which is important for cubesats \citep{lemmer2017propulsion} and is required for some multi-spacecraft missions that depend on precision flight and stability like the LISA laser interferometer mission \citep{ziemer2004microthrust} or the Habitable Worlds Observatory \citep{liu2014detection}. 

    For optimally efficient thrust, pure ion emission of a single mass species is desired \citep{hunter1960theoretical}. In practice, however, ionic liquid sources tend to fire a variety of compositional species at differing masses, with monomers, dimers, trimers, and higher mass clusters being created in the electrospray plume. A monomer is the bare ionic liquid cation or anion, and its emission yields the most mass-efficient thrust of all the ion species that can be evaporated from the emitter tip (that is, it achieves the highest $\Delta V$ per unit mass and thus can move a spacecraft the most per unit kilogram of propellant). Monomers attached to neutral anion-cation pairs to have a single net charge are referred to in the electrospray thruster community as dimers, and two neutrals with a monomer ion are referred to as trimers. Compared to monomers, such species have a higher per-molecule thrust (because they have a higher mass), but a lower mass efficiency (because the attached neutrals carry along mass that does not contribute to electrostatic acceleration). Higher order clusters and droplets are also observed, carrying away significantly greater amounts of propellant mass at lower effective velocity. This lowers the mass efficiency of the propellant and thus the maximum amount of spacecraft $\Delta V$ that can be achieved per unit kilogram of propellant, but it increases the thrust-to-power ratio \citep{lozano2005efficiency}.

    In the early 2000s, studies of RTIL electrosprays suggested that by limiting the flow rate of propellant to the emitter tip, emission in the pure ion regime (PIR) is possible \citep{gamero2001electrospray,romero2003source}. In PIR, ions are thought to be emitted from the liquid surface directly via field-assisted evaporation with minimal droplet emission \citep{lozano2005efficiency}. With the suppression of droplets and high mass clusters, PIR emitters are nominally capable of achieving orders of magnitude greater spacecraft $\Delta V$ over traditional thrusters per unit kilogram of propellant mass \citep{krejci2018space}. 
    
    For the first time in 2003, using a traditional capillary emitter, \citet{romero2003source} observed the PIR experimentally at the lowest flow rates (5.6e-13 kg/s), which was explained as the direct evaporation of ions as predicted in 1976 by \citet{Iribarne1976}. \citet{lozano2005efficiency} showed that an externally-wetted needle configuration could produce a higher hydraulic impedance and therefore more easily access the pure ion regime emission mode. It is worth highlighting, in the context of results to be shown in the present study, that the methodology for the capillary-based \citet{romero2003source} investigation accounted for the composition of ions and droplets collected from the entire beam. \citet{larriba2007monoenergetic} similarly performed a whole-plume study of a PIR plume, confirming its existence. It is otherwise both a common and practical approach to sample only a portion of the beam (as little as $<$ 1\%), with the emitting needle aligned coaxially with the downstream detection electronics \citep{lozano2005efficiency, miller2020measurement}. 

\begin{figure*}[h!]
        \centering
        \includegraphics[width=\textwidth, scale=1]{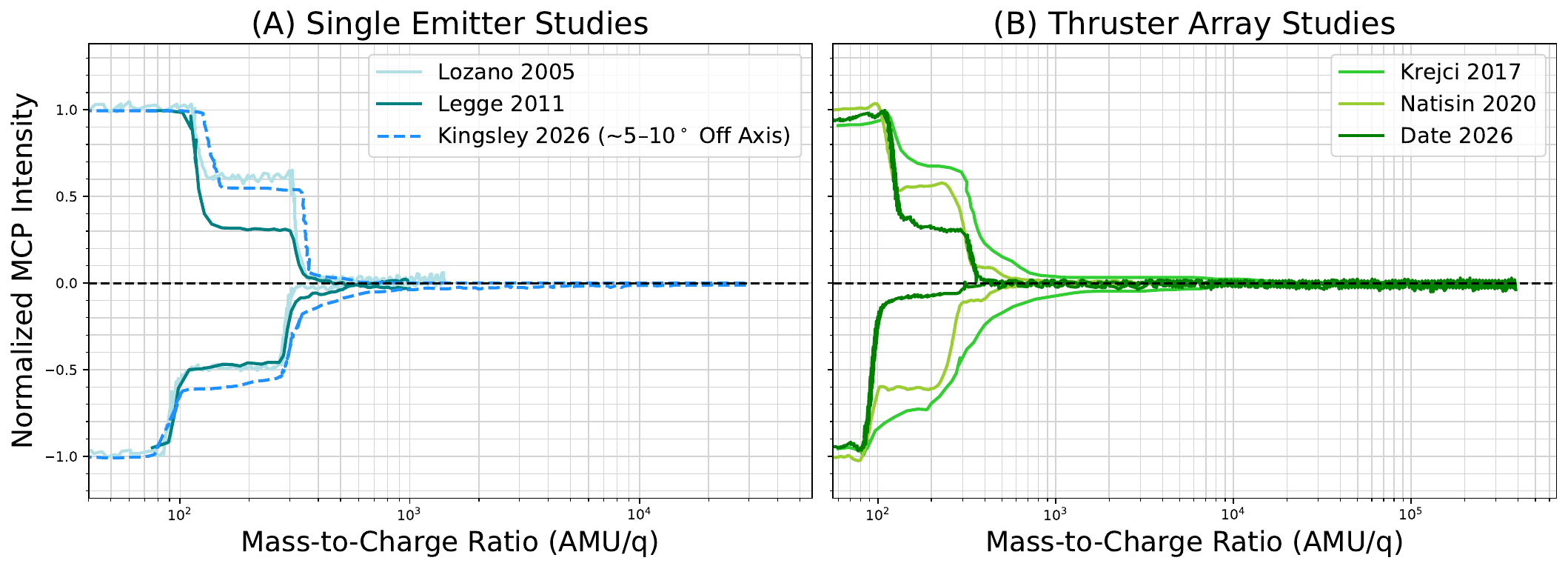}
        \caption{Examples of published time-of-flight data for (A) single emitters and (B) thruster arrays that suggest operation in the pure ion regime (PIR).}
    \label{PIRTOF} 
    \end{figure*}

    Recent years have seen an expansion in modeling and experimental work by several groups to further understand and improve electrospray thruster performance in the PIR \citep{gallud2022emission,smith2024propagating,asher2022multi,magnani2025modelling,takagiFeedPressure2025,lakshmi2026impact}.  Fig. \ref{PIRTOF} shows published time-of-flight curves from ionic liquid (A) single emitters \citep{lozano2005ionic,Legge2011porousmetals,Kingsley2026TPP} and (B) array thruster devices \citep{krejci2017emission, Natisin2020AFET, date2026electrospray} thought to be operating in or near the PIR, where each voltage step change corresponds to a population of ion species at the monomer and dimer masses. Flat regions between these steps indicate minimal fragmentation, and a flat, zero-voltage signal at high masses indicate minimal high mass or droplet content. 

    Nonetheless, despite the development of high efficiency single-emitter demonstrations, electrospray thruster arrays often fail to reach their anticipated mass efficiency with linear scaling applied for the number of emitters. Carefully tuned experiments measuring propellant mass flow rate and thruster $\Delta V$ output show discrepancies ranging from 23.1\% to 64.7\%, with significant portions of this ``missing'' propellant mass evidently not producing thrust or producing it at significantly lower amplitude than predicted \citep{natisin2021efficiency,de2025comparison}. Thus, the inability of modern electrospray EP devices to reach their theoretical potential may indicate incomplete understanding of the underlying physics that govern the electrospray process, the downstream plume dynamics, the measurement techniques themselves, or some combination thereof.

    The missing mass that has prevented electrospray propulsion from reaching its full theoretical potential may in part be the result of trimers and additional heavy mass clusters undergoing prompt fragmentation in a region indiscernibly close to the emission meniscus, as this complicates ion flight time and energy measurements \citep{Smith_2025}. However, for a beam composed entirely of emitted monomers, dimers, and trimers, prompt fragmentation likely cannot account for all of the observed missing mass \citep{Smith_2025}. Differences in angularly-dependent plume composition may help shed light on this phenomenon and provide critical experimental data for modeling efforts, especially with an eye towards any ability to characterize the energy distribution of neutrals and high mass molecules present in the plume \citep{naemura2025direct,bendimerad2024investigating,tahsin2025cross}.

    Further, integrated propulsion devices are observed to be lifetime-limited due to electrical shorting, thought to be the result of propellant accumulation from off-axis, angular emission onto source geometries and surfaces \citep{thuppul2021mass,krejci2017emission}. Imperfections in the emitter geometry such as micron-scale manufacturing defects on the emitter tip and/or emitter placement error \citep{smith2026kinetic,takagi2024simple} and the existence of multiple emission sites \citep{takagi2025variability,kingsleyTPPEmitter2025} may lead to significant off-axis emission and spatial non-uniformity \citep{asher2022multi,guerra2016spatial}. Furthermore, clusters experience collisional or temperature-induced fragmentation, leading to speciation within the plume itself and resultant changes in axial composition and thruster performance \citep{miller2020measurement,bendimerad2024investigating,tahsin2025cross}. Emitter-scale processes that contribute to this accumulation are often exacerbated on the thruster scale due to hundreds of emitters operating in tandem. Thus, a better understanding of the angular firing properties of the ejecta plumes, especially with respect to composition and the presence of neutrals, is important to the long-term realization of the full potential of these EP sources.

    Time-of-flight (TOF) mass spectrometry (MS) detectors are commonly used to measure the composition of the plumes, typically by quantifying the time-dependent voltage rise of a detector, usually a microchannel plate (MCP) or channel electron multiplier (CEM) detector, behind an electrostatic gate. Historically, despite meticulous alignment of the emitter needle axis with the detector flight path, it has not been practical or presumed essential to actuate the ion plume itself with respect to the downstream detector; results capturing a small patch of the emission plume along a single sampled path are often used to extrapolate whole-plume composition. With thruster arrays composed of many emitters, it has generally been assumed that TOF measurements provide average emission characteristics of the entire system. However, depending on the geometry of the TOF gate and detector, a TOF instrument could be sampling between as little as $<$ 0.01\% (Ø 0.5 cm detector, 116 cm from source as in \citet{petro2020characterization}) and as much as 30\% (Ø 8 cm detector, 70 cm from source as in \citet{natisin2021efficiency}) of the integrated plume. The high divergence measured in arrays ranging from half angles of 45 \citep{petro2020characterization} to 75 degree \citep{natisin2021efficiency} are not explainable by space-charge alone \citep{smith2024propagating,jia2026fast}, further suggesting that off-axis emission could be common across an array.

    \subsection{Previous Studies of Angular Structure}

    Chiu \textit{et al.} conducted some early two dimensional studies of strips of externally-wetted emitter tips to provide qualitative, constrained two-dimensional compositional studies of 1-ethyl-3-methylimidazolium bis(trifluromethylsulfonyl)imide (EMI-Im) \citep{ticknor2009mass,chiu2005mass,chiu2007vacuum}. The emitter ribbon was placed on a one-dimensional rotation stage, and vacuum was broken and the strip was rotated 90 degrees to analyze the second dimension. Those studies found mixed droplet and ion emission, with higher operating voltages creating wider plumes with greater droplet current and higher average droplet size.  While this important and pioneering study provided valuable and unique data, the system could not study both dimensions simultaneously, and thus it could not study regions of the plume that were off of both emitter axes. Furthermore, the plume dynamics themselves could change during the chamber pressure cycling, reconfiguration, and pumpdown cycle, and much of the compositional data is presented qualitatively rather than quantitatively.

    Perez-Martinez et al. used an MCP to amplify emitted EMI-BF$_4$ impacting a phosphor screen to measure plume intensity in two dimensions \citep{perez2011ionic,perez2012visualization}. While the system could not in general quantify compositional variation, the use of retarding potentials and electrostatic deflection enabled the detection of an energetic beam of neutral molecules within the plume. Similarly, Naemura and Takao used deflection plates to directly measure neutral density of EMI-BF$_4$ and 1-ethyl-3-methylimidazolium dicyanamide (EMI-DCA) with a CEM \citep{naemura2025direct}, though 2D spatial data was not possible.
    
    Other groups have studied angular properties with a single axis of rotation of the source itself or with translation of the downstream diagnostic instruments, which sometimes enabled two-dimensional plume measurements \citep{miller2014electrospray}. Wirz \textit{et al.} developed the HOAGIE apparatus, which uses a two-dimensional translation of downstream instruments to provide planar sweeps of plume current densities of capillary emitters \citep{thuppul2021mass,collins2022high,uchizono2020emission}. Those studies observed EMI-Im plume emission tilts of up to 15$^\circ$ at higher acceleration potentials \citep{uchizono2020emission}, while models found an increased density of high mass droplets in the plume centers. Molecular composition as a function of angle is not directly measurable from these results.

    \citet{perez2026quasi} performed a careful full-beam characterization of negative-mode capillary emitters using four EMI-based ionic liquids with 30-36 $\mu$m tip diameters. Their apparatus used six concentric annular ring collectors perpendicular to the spray axis, together collecting approximately 50\% of emitted current directly, with the remainder accounted for after transparency corrections. A quasi-ionic (QI) regime was identified in which droplet current contributed as little as 0.5-3\% of total current, yet represented a substantial fraction of total emitted mass.

    \citet{schroeder2023angular} studied EMI-BF$_4$ sprayed from a porous emitter with a CEM-based TOF system. A rotational stage that allowed the beam angle to be varied in one dimension was used to produce a simulated full-beam TOF curve, from which the relative compositions of the monomers, dimers, and trimers were measured as a function of accelerating voltage rather than angle. Using a Kalman update algorithm presented in \citet{jia2022quantification}, \citet{jia2024numerical} analyzed an 800 V subset from \citet{schroeder2023angular} to provide true angular-dependent compositional measurements. This study found that dimers dominated the center of the beam, monomers dominated the extremities, and that trimers and high-mass species were most prevalent at the center but weakly present everywhere.

    However, because \citet{schroeder2023angular} study could only vary one-dimension, there is no guarantee that the studied beam slice ran along the true center line of the emission beam (that is, through the beam's peak amplitude). A key finding of the original study was that the beam's center angle varied from +7$^{\circ}$  to -8$^{\circ}$ across an acceleration voltage range 800 to 950V. Indeed, a second emission site found at high potential in \citet{schroeder2023angular} was weakly identified and posited to be ``off axis in the direction orthogonal to that investigated''. Regardless of the slice's exact position in the full two-dimensional beam path, extrapolation of the results across one dimension to a full plume composition necessarily assumes azimuthal symmetry of the plume which by definition could not be verified without analysis in the second dimension. 

    We have previously reported a new method of two-dimensional steering of vacuum electrospray plumes \citep{ulibarri2025direct}, where a dual-axis goniometer is used to pivot an emitter and its extractor about the emission site to enable study of the plume as a function both $X$ and $Y$ angles with fixed downstream detectors. Those preliminary results showed significant angular variation in the compositional content of the plume, in particular with heavy molecules comprising a larger share of the total current in the center and at the extremities. However, the difficulty of controlling the device manually meant that we were only able to provide a one-dimensional scan of the plume, with the second dimension only being used to confirm that the scan axis went through the center of the plume. The one-dimensional nature meant that those results perforce assumed azimuthal symmetry, as in previous studies. Further, only one scan was taken, meaning that statistical analysis of the angular profile was impossible, with somewhat arbitrary proxy error bars provided only by estimation. The precise emitter geometry was not known, inducing greater uncertainties for any attempt to replicate the data through modeling. Lastly, that study was limited to masses below 2800 AMU/q, and there was at that time no attempt to characterize energetic neutrals within the plume \citep{ulibarri2025direct}.

    \subsection{The Present Work}

\begin{figure*}[h!]
    \centering
    \includegraphics[width=\textwidth, scale=1]{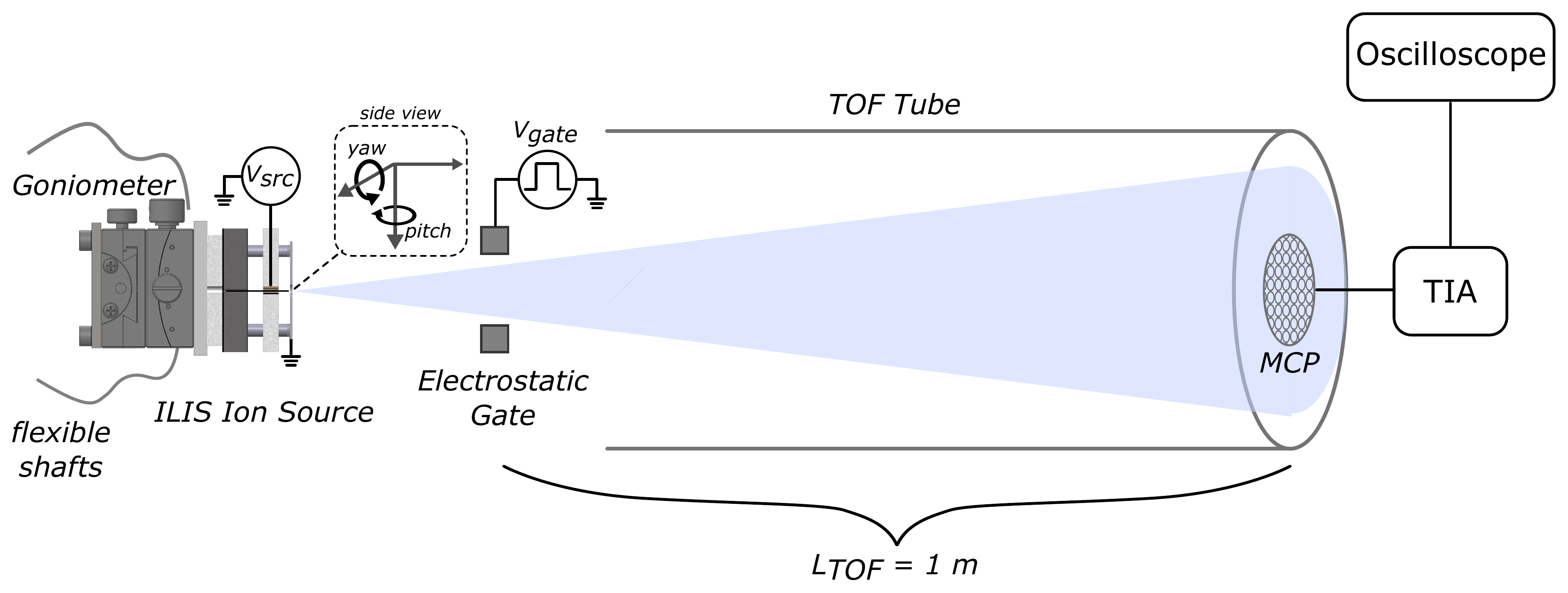}
    \caption{Diagram of the experimental apparatus for the two-dimensional compositional plume studies.  
    }
    \label{fig: experimental diagram} 
    \end{figure*}

    Here we present a full two-dimensional, 2$^\circ$ resolution compositional characterization of a vacuum electrospray plume generated by a 9.6 $\mu$m radius tungsten needle externally-wetted by the ionic liquid EMI-BF$_4$. Fig. \ref{fig: experimental diagram} shows a diagram of the experimental setup.  A motorized, computer-controlled setup for the dual-axis goniometer was used to take multiple TOF acquisitions at each angular cell, and repeated measurements out to $>13,500$ AMU/q enabled further statistical analysis and high mass droplet detection. The precise emitter geometry is well-characterized through the use of a laser profilometer that generated a 3D model of the emitter needle's precise location with respect to the extractor.

    We find significant azimuthal asymmetry in the data, both in total plume current and within individual species data, as different compositional species are found to peak in different parts of the plume. Based on averaged radial scans of the plume, heavy molecules peak at the plume center while monomers form a ring structure around it, both consistent with the previous study. However, the heavy particles are observed to effectively form a distinct sub-plume with a flight path that deviates from the main beam by 1.5$^\circ$. These results illustrate the importance of sampling electrospray plumes in two dimensions and, combined with prior evidence, definitively establish angular compositional heterogeneity of ionic liquid electrospray plumes.

    We further find that there exists significant two-dimensional variation in the baseline detector voltage (the signal floor upon which all other observed features sit), strongly suggesting that energetic neutrals represent an important fraction of the overall plume content. The energetic neutrals likewise form a sub-plume observed to deviate from the primary beam by 3.4 $^\circ$. This demonstrates an improved method for characterizing the spatial distribution of such neutrals, which has heretofore been limited despite their importance for the efficiency of the propulsion systems \citep{naemura2025direct,geiger2024qcm,perez2012visualization}.

   The plume analyzed in this study is consistent with a cone-jet emission site rather than pure ion emission, but if the plume is sampled sufficiently far off-axis, it appears as though it is a highly efficient PIR plume. The TOF-estimated $\Delta V$ propulsive efficiency varies by a factor of 6 across the plume, meaning that it would be quite easy to mistake cone-jet operation for PIR emission if whole-plume composition is not directly measured. In such a case, a single sampling axis could incorrectly point towards a preponderance of ions and a lack of droplets and high mass clusters. This would provide a simple and straightforward explanation for part of the missing mass question in electrospray propulsion.


\section{Experimental Design} \label{sec: experimentaldesign}

    \subsection{Two Dimensional Control and TOF Setup}
    Details of the goniometer setup and $X$-axis assignment are given in \citet{ulibarri2025direct}, and details of the TOF setup are given in \citet{cogan2023electrospray}, and thus they are only described briefly here. A dual-stage Opto-Sigma GOH-40B35 goniometer is used in-vacuo to directly steer a tungsten needle and its attached extractor in two dimensions rotating about the tip of the needle with $20^\circ$ of throw range in the $X$ axis (referred to as yaw in the previous study) and 15$^\circ$ in the $Y$ axis (pitch in the previous study). The goniometer is controlled by flexible shafts attached to rotational feedthroughs on the vacuum chamber. An electrostatic gate is composed of two electrodes which are held at $\pm$ 1kV to deflect ions away from a 1 meter flight tube ($L_{TOF}$) and a Hamamatsu F1217-011 MCP detector with a transimpedance amplifier (TIA) at its termination. The increased signal-to-noise ratio of the two dimensional setup often overpowers the range of the TIA, and thus, as in previous studies, we used an MCP setting of 900 V rather than the nominal 1000 V. The electrodes are rapidly pulsed to ground, suddenly allowing the particles to reach the detector, thus creating a rising TOF curve. The beam characterized in the present study has a half-angle of about $15^\circ$, which for the 42 mm diameter MCP at a 1 meter flight distance corresponds to approximately 1.6\% of the plume being incident on the MCP at any given angular cell.

    Whereas previous work required manual goniometer actuation and data logging, the present work features a fully-automated data scanning process. The scan starts at the lowest selected negative angles in both $X$ and $Y$, and then scans in fixed steps across $X$ to the highest selected angle. It then increments in $Y$ before scanning again from the lowest $X$ angle to the highest. Emitter current and voltage from the Keithley 2657A source meter unit (SMU) were averaged over the time the system was paused at each cell.

    \subsection{Experiment}

    A two-dimensional quantitative plume analysis experiment was performed with two motorized scans, henceforth called Runs A and B, taken sequentially. The needle was conditioned by firing for 1 hour before Run A, and both runs took 38 minutes. Runs A and B were separated by roughly one hour of continuous source operation. Scans were taken at 2$^\circ$ resolution in both dimensions. The electrostatic gate was pulsed at 500 Hz, and two oscilloscope spectra acquisitions, each composed of 100 averaged gate pulse triggers, were taken for each cell.  A time domain window sufficient to capture mass data out to $>13,500$ AMU was used. Both runs were conducted with a source acceleration voltage of 1450V. Run A had an average emitter current of 480 $\pm$ 20 nA while Run B had an average of 470 $\pm$ 20 nA.

    Supplemental Fig. \ref{fig: run910profilometry} shows source profiles from the Keyence VK-X260 laser profilometer, including the $X$ and $Y$ profile scans across the needle's center, taken immediately after Run B. Scanning electron microscope images of the needle tip are shown in Supplemental Fig. \ref{fig: B4SEM}. The needle's radius of curvature was measured to be $\sim$9.6$\mu m$, and the needle tip was offset 50~$\mu$m laterally (left) and 70~$\mu$m vertically (above) from the center of the 2.5~mm diameter extractor aperture. The tip was recessed 600~$\mu$m below the top face of the 1.5~mm thick extractor plate.


    \subsection{Beam Composition Mapping}\label{yield}
    Compositional species yields were calculated in the same manner detailed in \citet{ulibarri2025direct}, specifically that study's Fig. 2, with one important change. Whereas in the previous study the background MCP voltage floor (henceforth baseline) was treated as a confounding factor and eliminated through subtraction, here we recognize it as an important fundamental feature of the emission that characterizes energetic neutrals which pass through the electrostatic gate regardless of its deflection potential. As before, the voltage rise from the upper trimer bound to an arbitrary point called the high mass cutoff at the end of the recorded data is taken as the heavy particle or droplet yield. The high mass cutoff is now 13,000 AMU (as opposed to 2800 AMU in the previous study) for the two dimensional TOF. These yields were then used to create species-specific heatmaps that show the plume structure. Further details of the compositional yield measurements are given in the Appendix Sec. \ref{supp_mol}.


    \subsection{Radial Statistical Analysis} \label{statistics}

\begin{figure*}[h!]
    \centering
    \includegraphics[width=\textwidth, scale=1]{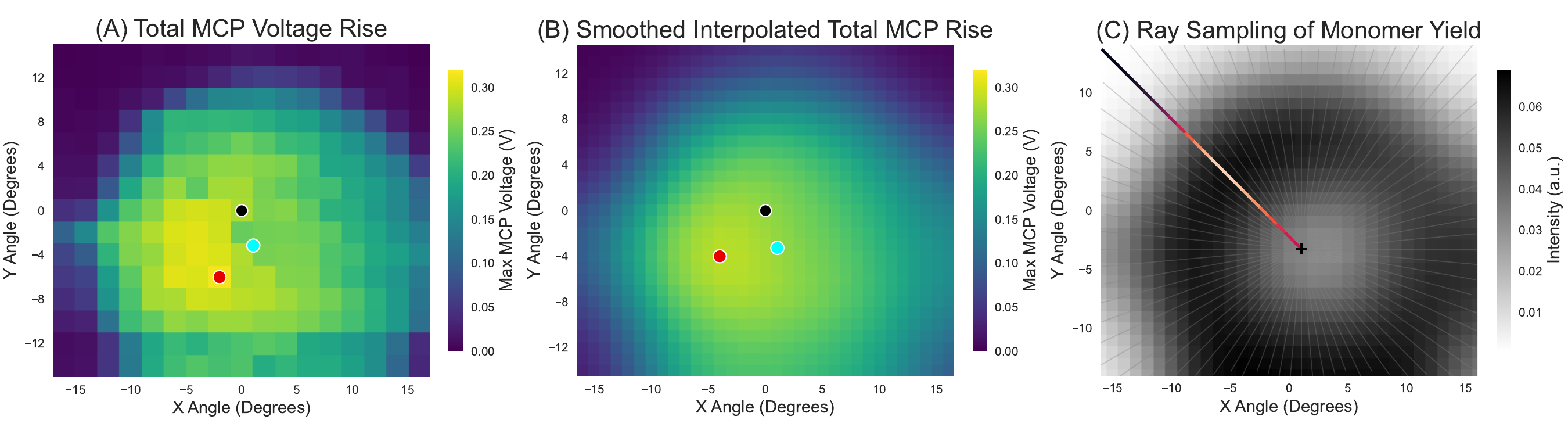}
    \caption{An overview of the radial statistical analysis using data from Run A as an example, as described in Sec. \ref{statistics}. (A) shows the raw MCP signal as a function of goniometer angle in the original 2$^\circ$ resolution. A black dot shows the (0,0) location, which is defined as the emitter needle's axis. A red dot shows the signal's peak amplitude location, and a cyan dot shows the signal center found by a 0.35 threshold centroid calculation. (B) Shows the same data after two dimensional Gaussian smoothing and interpolation to 1$^\circ$ resolution. The red and cyan dots mark the peak and centroid centers calculated from this processed data. (C) shows the monomer ion yield in a black and white intensity map with the 72 evenly spaced azimuthal rays used for analysis. One ray, at 140$^\circ$, is shown with a color intensity map to illustrate the values along that reinterpolated ray measured for analysis.}
    \label{interpolation} 
    \end{figure*}

    To determine the plume center, two-dimensional Gaussian smoothing was applied to the MCP intensity map with a $\sigma$ of 1.5$^\circ$, and the smoothed two-degree resolution data was then interpolated to one-degree resolution. The center of the resultant intensity map was then calculated by finding the centroid of the data above a 35\% threshold, required because some scans do not include the full plume, artificially pulling the centroid away from the missing data. Fig. \ref{interpolation} (A) shows a raw two degree resolution MCP voltage rise heatmap from Run A. Fig. \ref{interpolation} (B) shows the same data smoothed and interpolated to one degree resolution. Both plots have a black dot at the on-axis needle zero location in $X$ and Y, a red dot at the peak intensity, and a cyan dot at the centroid-defined center.

    The species-specific heatmaps described in \ref{yield} were smoothed and interpolated in the same manner as the MCP signal. Using the MCP signal centroid as the plume center, radial intensity profiles were then extracted to determine the intensity of each species as a function of radius.  72 slices were then taken across the plume to provide 5$^\circ$ separated ray lines for analysis, as shown in Fig. \ref{interpolation} (C), which uses the monomer yield as an example species. The 72 rays selected for analysis emanate out from the MCP centroid center, with a different color map on one ray at 140$^\circ$ to show the interpolated data that was selected for study. Intensity was then sampled at 1$^\circ$ steps in radial distance from the MCP centroid center, calculated via bilinear reinterpolation of the smoothed, 1$^\circ$ interpolated cartesian heatmap. Runs A and B were treated as independent measurements with distinct plume centers.

    The statistical error bar methodology is described in detail in the Appendix Sec. \ref{error}; briefly, the error bars represent the standard error of means combining azimuthal variation across the 72 radial rays and measurement variation across Runs A and B, added in quadrature, with an effective sample size $n_\mathrm{eff}$ used to conservatively account for auto-correlation between adjacent azimuthal samples \citep{bayley1946effective}.


    \section{Results} \label{sec: results}

    \subsection{2D Species Yields}

    \begin{figure*}[h!]
    \centering
    \includegraphics[width=\textwidth, scale=1]{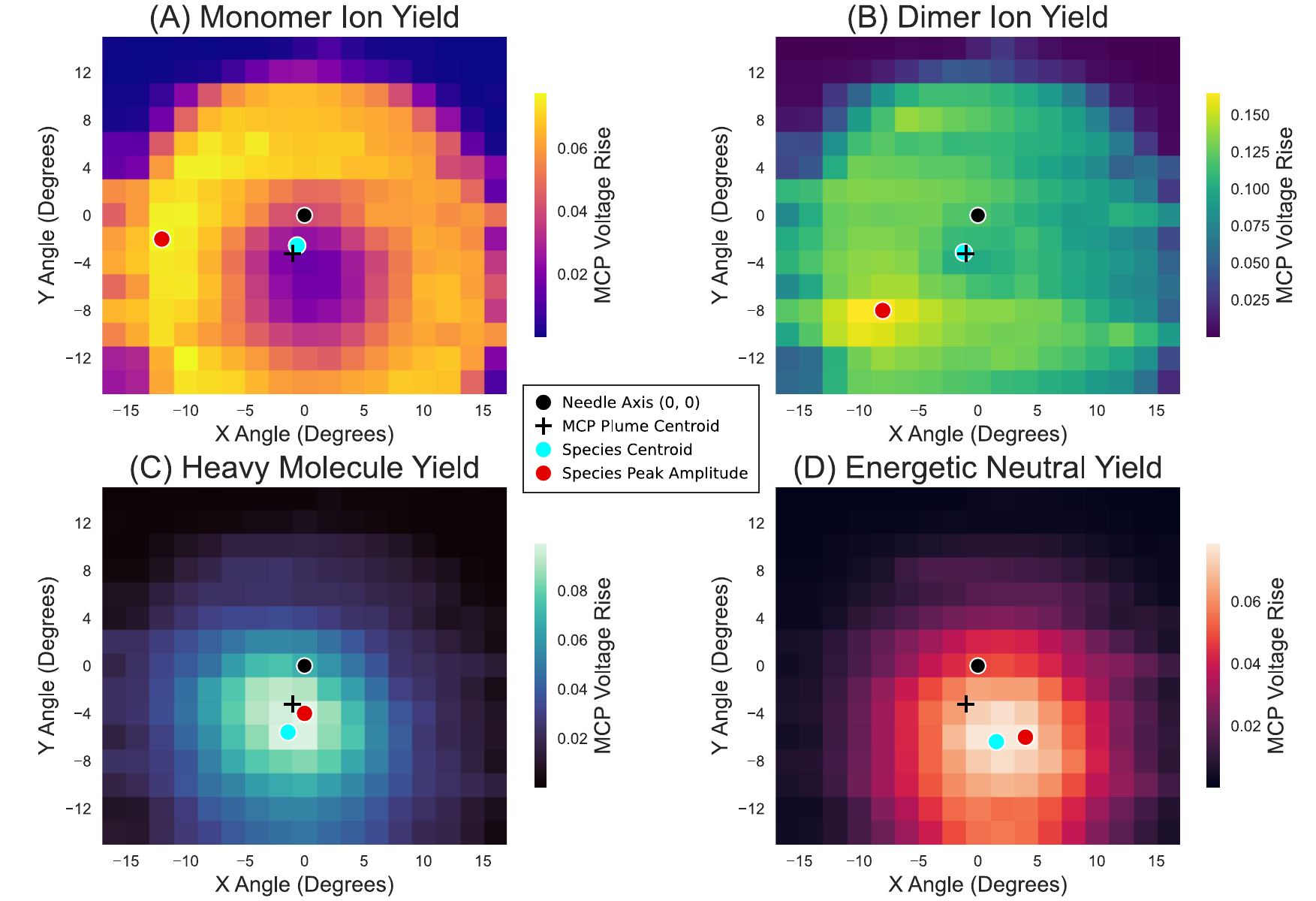}
    \caption{Heatmaps of the major species yields of Run A. Black dots show (0,0), defined by the needle axis, while black ``+'' signs mark the 35\% threshold centroid of the total plume current (the raw MCP voltage rise of all combined species). As these two black markers denote constants for each run, neither moves position between the four subplots. Red dots show the peak amplitude location for the subplot species, and cyan dots denote the 35\% threshold centroid center of the subplot species. The emitted current is shown in the Supplemental Fig. \ref{output}, from which it can be seen that the asymmetries observed in these heatmaps cannot be attributed to erratic source behavior. The corresponding heatmap series for Run B is shown in the Supplemental Fig. \ref{R2A}.}
    \label{heatmaps}
    \end{figure*}

    Heatmaps of the raw species yield for Run A are shown in Fig. \ref{heatmaps}, and corresponding data for Run B is given in the Supplemental Fig. \ref{R2B}. The same data with a 2D Gaussian smoothing, used for the radial statistics, are shown in Supplemental Figs. \ref{R2A-smoothed} and \ref{R2B-smoothed}. The results are summarized in Table \ref{TableAngles}, which gives the emission currents and angles for each of the two experimental Runs. The black dot in Fig. \ref{heatmaps} represents the needle axis (0,0) location in $X$ and $Y$. The total MCP signal centroid is marked by a black cross. As both of these black markers are not species-dependent, they are fixed in place and do not move from subplot-to-subplot. The red dots mark the location of the max signal intensity for each species, and the thresholded centroid of each species is marked by cyan dots.
    
    It can be seen visually that there are nontrivial asymmetries in  plume composition, with the heavy species yield being shifted down from the MCP centroid an average of 1.5$^\circ$ while the energetic neutral yield is shifted an average of 3.4$^\circ$ down and to the right. The dimers are well-centered, but their peak intensity is 8.6$^\circ$ down and to the left in Run A and 5.8$^\circ$ in Run B. The left side of the monomer yield is noticeably brighter than the right side. Supplemental Fig. \ref{output} shows the output currents from the SMU, from which it can be seen that the emitter source was largely stable during the experiment. Thus, the asymmetries in the data cannot be attributed to erratic source behaviour.

\begin{table*}
\centering
\renewcommand{\arraystretch}{2.0}
\setlength{\tabcolsep}{4pt}
\footnotesize
\resizebox{\textwidth}{!}{%
\begin{tabular}{>{\centering\arraybackslash}m{1.4cm} !{\color{lightgrey}\vrule} >{\centering\arraybackslash}m{1.6cm} !{\color{lightgrey}\vrule} >{\centering\arraybackslash}m{1.8cm} !{\color{lightgrey}\vrule} >{\centering\arraybackslash}m{1.9cm} !{\color{lightgrey}\vrule} >{\centering\arraybackslash}m{1.9cm} !{\color{lightgrey}\vrule} >{\centering\arraybackslash}m{1.9cm} !{\color{lightgrey}\vrule} >{\centering\arraybackslash}m{1.9cm}}
    \toprule
    \multicolumn{7}{c}{\cellcolor{xkcdvioletblue}\textbf{\small\textcolor{white}{Plume Angular Properties}}} \\
    \multicolumn{7}{c}{\textbf{\small Ø 2.5 mm Extractor, Emitter Displacement 50 $\mu$m (X-axis), 70 $\mu$m (Y-axis)}} \\
    \arrayrulecolor{lightgrey}\midrule\arrayrulecolor{black}
    \textbf{\small Run} &
    \textbf{\small\shortstack{Emission\\Current}} &
    \textbf{\small\shortstack{Plume\\Angle ($^\circ$)\\[2pt]$(x_0, y_0)$}} &
    \textbf{\small\shortstack{Monomer\\Angle ($^\circ$)}} &
    \textbf{\small\shortstack{Dimer\\Angle ($^\circ$)}} &
    \textbf{\small\shortstack{Heavies\\Angle ($^\circ$)}} &
    \textbf{\small\shortstack{Neutral\\Angle ($^\circ$)}} \\
    \midrule
    \textbf{\small Run A} &
    $480 \pm 20$ nA &
    $(-1.0, -3.3)$ &
    \shortstack{$(x_0+0.5,$\\$y_0+1.0)$} &
    \shortstack{$(x_0-0.2,$\\$y_0+0.3)$} &
    \shortstack{$(x_0-0.5,$\\$y_0-2.0)$} &
    \shortstack{$(x_0+2.7,$\\$y_0-2.7)$} \\
    \arrayrulecolor{lightgrey}\hline\arrayrulecolor{black}
    \textbf{\small Run B} &
    $470 \pm 20$ nA &
    $(1.0, -3.3)$ &
    \shortstack{$(x_0+0.3,$\\$y_0+0.5)$} &
    \shortstack{$(x_0-0.3,$\\$y_0+0.1)$} &
    \shortstack{$(x_0-0.5,$\\$y_0-0.8)$} &
    \shortstack{$(x_0+2.5,$\\$y_0-1.8)$} \\
    \midrule
    \textbf{\small Polar Averages} &
    $475 \pm 14$ nA &
    \shortstack{$(R_0, \phi_0)$\\$= (3.4, 270^\circ)$} &
    \shortstack{$\Delta$ from\\centroid:\\$(0.9, 61^\circ)$} &
    \shortstack{$\Delta$ from\\centroid:\\$(0.3, 142^\circ)$} &
    \shortstack{$\Delta$ from\\centroid:\\$(1.5, 247^\circ)$} &
    \shortstack{$\Delta$ from\\centroid:\\$(3.4, 320^\circ)$} \\
    \bottomrule
\end{tabular}%
}
\caption{Summary of experimental runs showing average emission current and measured plume angles for each species. Monomer, dimer, heavy, and neutral angles are expressed relative to the MCP centroid plume angle, defined as $(x_0, y_0)$. The monomer and dimer ion species are well-centered, featuring average offsets of less than one degree from the MCP centroid, but the heavy species and neutrals exhibit significant divergence from the overall beam center, averaging $1.5^\circ$ and $3.4^\circ$ offsets respectively.}
\label{TableAngles}
\end{table*}

    \subsection{Radial Species Yields}

\begin{figure*}[h!]
    \centering
    \includegraphics[width=\textwidth, scale=1]{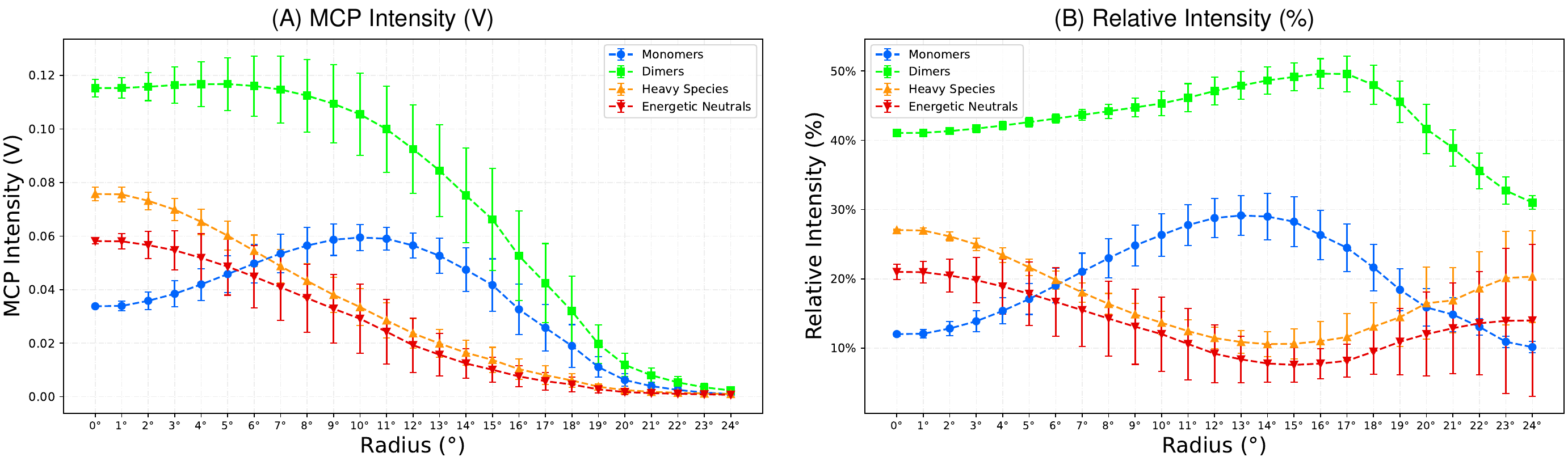}
    \caption{The yields of each compositional species as a function of radius from the plume center (not the needle axis). Dimers form the majority of ion yield and have a flat distribution out to about $10^\circ$. Monomers are comparatively weak at the center, with only slightly more than half their absolute intensity max, which is found at about 10$^\circ$ off center. Heavy molecules and energetic neutrals are at a maximum at the center, but both form a significant portion of the plume.}
    \label{radial} 
    \end{figure*}

    The results from the the radial statistical analysis are shown in Fig. \ref{radial}. Plot (A) shows the absolute intensity of each species in MCP voltage as a function of radius, while plot (B) shows the relative intensity (wherein all species are reported as a percentage of the total plume intensity). The monomers show a relative minima at the beam center, rising to maximum intensity at about 10$^\circ$ off-center. However, as a percentage of the plume, monomers reach a maximum of 29\% at 13$^\circ$ off-axis, up from the 12\% observed at the center. The dimers exhibit a flat intensity out to about 6$^\circ$, where they begin to drop off. The relative dimer contribution rises slightly (from 40 to 43\%) over this angular space, driven by the decreasing contributions from heavy species and energetic neutrals, which contribute 27\% and 21\% respectively at the beam center. These heavy species and neutrals drop off to about 10\% at 15$^\circ$, after which their relative contributions rise again at the extremities of the plume.

\section{Discussion} \label{sec:discussion}

 \subsection{Specific Impulse} \label{isp}

  The allure of PIR lies in that ions possess much higher velocities and thus greater specific impulse ($I_{SP}$) compared to high mass-to-charge ratio droplets when accelerated through identical potential differences. Spacecraft $\Delta V$ per unit mass of propellant scales directly with $I_{SP}$, making it a key metric for propulsive efficiency. Reported PIR electrospray systems achieve specific impulses of $\sim$3000 s, comparable to the most efficient EP systems like gridded ion thrusters \citep{petro2017survey}, whereas droplet modes may only achieve $\sim$100-400 s of $I_{SP}$ \citep{caballero2026JFMprepreint}. 

    The `tyranny of the rocket equation' dictates that such a reduction in $I_{sp}$ severely limits the resultant $\Delta V$ of the propulsion source. At the level of 3000 s $I_{sp}$, a mission with a 5 km/s $\Delta V$ requirement would require 16\% of the spacecraft mass to be propellant. At the 100 second $I_{sp}$ level where capillaries operate, the spacecraft would then require nearly 99\% of the total mission mass to be propellant. Therefore, it is paramount to fully capture the correct TOF window when deriving species performance parameters and comparing TOF results to analytical balance or thrust stand measurements.

  \begin{figure*}[h!]
    \centering
    \includegraphics[width=1\textwidth]{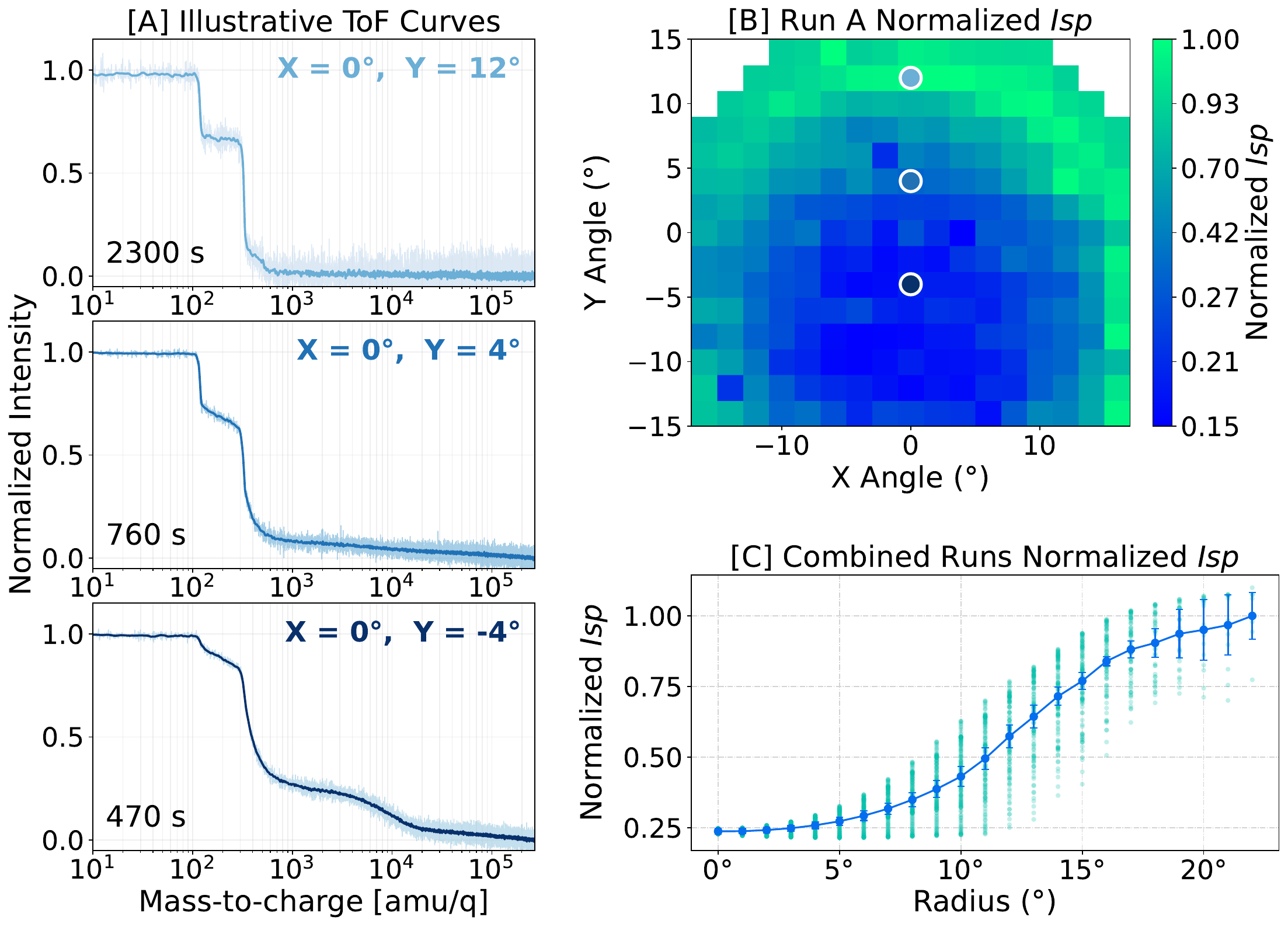}
   \caption{[A] Representative Run A TOF spectra at three angular positions (white circles in [B]), demonstrating the shift from 
    ion-dominated emission at high off-axis angles to droplet-dominated emission near 
    centerline. [B] Normalized heat map of TOF-derived $I_{sp}$ as a function of plume 
    angle in Run A. [C] The combined normalized $I_{sp}$ radial measurements from both Runs.}
    \label{fig: ispheatmap}
    \end{figure*}
  
  Specific impulse can be derived from the TOF curve by calculating mass flow rate ($\dot{m}$) and thrust ($F_{T}$) by integrating the normalized time-domain TOF curve and scaling by the emitted current via,

     \begin{equation}
        \dot{m}_{\text{TOF}} = \frac{4V_{0}}{L_{\text{TOF}}^2} \int_0^{\infty} I(t)\, t\, dt
    \end{equation}
    
    \begin{equation}
    F_{T, TOF} = \frac{2V_{0}}{L_{TOF}} \int_0^{\infty} I(t)\, dt
    \end{equation}

    \begin{equation}
    I_{sp} = \frac{1}{g} \frac{F_{T}}{\dot{m}}
    \end{equation}

    \noindent where the TOF subscript represents the method of calculation, $g$ is the gravitational acceleration and $I(t)$ is the normalized TOF current multiplied by the emitted current \citep{Natisin2020AFET, petro2020characterization}.

  Fig. \ref{fig: ispheatmap} shows the normalized heat map of the $I_{sp}$ measured for every cell of the Run A data. Three exemplar TOF curves, taken at 8 degree intervals from $y = -4^\circ$ (the beam center location) to $y = +12^\circ$ are shown at the left, with dots on the heat map to illustrate their cell locations. At the high-off-axis extremity, a highly flat TOF curve with minimal fragmentation and little or no observable droplet content is observed. This spectrum could easily be mistaken for PIR emission with a calculated $I_{sp}$ of 2300 s (ignoring sub-nominal MCP voltage calibration), were it not known to be an off-axis measurement. In contrast, the TOF curve at the beam center features significant fragmentation and higher mass clusters and droplets, with a calculated $I_{sp}$ of only 470 s, about 20\% of the maximum. However, because our MCP was operating at sub-nominal voltage, we cannot be exactly certain of the total collected current at any particular cell, meaning precise $I_{sp}$ quantification is not possible. We thus report Fig. \ref{fig: ispheatmap}'s heatmap in normalized terms to illustrate the larger point: whatever the peak $I_{sp}$ at the plume's edges, the plume's center instead exhibits as little as 15-20\%. The center region is largely consistent, with most values ranging between 18-22\% of the maximum. Thus, we have employed a nonlinear equalized color scale to maximize visual contrast in the densest region of the data distribution \cite{PIZER1987355}. The combined radial $I_{sp}$ measurements from both Runs range from 24\% in the center to a maximum at the edges of the plume.

\subsection{Two Dimensional Plume Structure}
    The plume structures clearly show cone-jet operation from this emitter source, and PIR emission very obviously cannot explain the heavy species and droplets observed near the beam center. However, Figs. \ref{heatmaps} and \ref{radial} both show that sampling different parts of the plume may yield quite different composition profiles. Given that the beam is observed to fire off-axis, and that this is not unique to our system, \citep{schroeder2023angular,uchizono2020emission}, it is possible that many published studies may be unknowingly sampling off-axis emission.

    To highlight exactly what this potential effect is with respect to electrospray propulsion systems, we consider the specific impulse as a function of angle for the present study. Because the specific impulse of the beam, and thus the spacecraft propulsive efficiency $\Delta V$, is fundamentally tied to the composition, such measurements necessarily depend on the specific part of the plume being analyzed.

\begin{figure*}[h!]
    \centering
    \includegraphics[width=\textwidth]{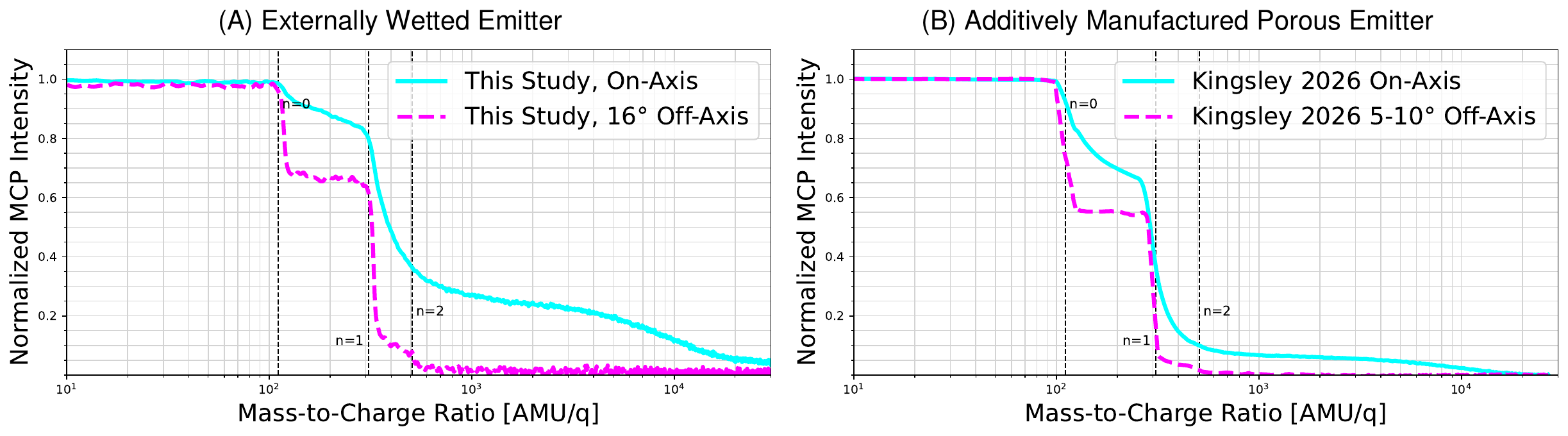}
    \caption{On- and off-axis TOF data from (A) the present study and (B) an additively manufactured porous emitter at $V_{acc} = $2500 V, from Ref. \citep{Kingsley2026TPP}. If only the off-axis data were sampled, either system would appear to be in highly efficient pure-ion-mode operation.}
    \label{fig: bryceangle}
    \end{figure*}

    Fig. \ref{fig: bryceangle} shows two pairs of TOF curves. (A) shows two curves from the present study's Run A, with one taken at the MCP centroid center and another taken 16$^\circ$ off-axis. The on-axis TOF curve exhibits significant fragmentation (as the regions between species bounds are not flat) and quite significant heavy particle and droplet content, yielding a specific impulse of just 470 s. By contrast, the off-axis curve features no meaningfully observable fragments, heavy species, or droplets, yielding a specific impulse of 2300 seconds. That is, without the ability to resolve the plume angle, these TOF measurements may underestimate the mass flow rate by a factor of 5, with a corresponding reduction in propulsive $\Delta V$. To show that this discrepancy is not merely a feature of this one, present study, or of externally-wetted emitters specifically, Fig. \ref{fig: bryceangle} (B) shows a similar pair of on- and off-axis measurements from \citet{Kingsley2026TPP}, which used an additively manufactured porous emitter loaded with EMI-BF$_4$ at a 2500 V acceleration potential. In that study, the on-plume-axis TOF curve yielded an $I_{SP}$ of $\sim$ 1800s, but measurements 5-10$^\circ$ off-axis were as high as $\sim$ 4600 s, presenting a 2.6 difference factor in flow rate and propulsive efficiency \cite{Kingsley2026TPP}.

    It is important to recall that minute defects in emitter geometries may yield significant plume divergence from the emitter axis \citep{whittaker2023electrospray,smith2026kinetic}. Theoretical results suggest that extractor-emitter misalignment has a measurable effect on the plume emission angle, but off-center emission sites, often thought to be the result of microscopic imperfections on the emitter surface, result in much more significant shifts \citep{smith2026kinetic}. Current manufacturing methods are incapable of producing emitters without sub-micron-scale defects, meaning that these defects alone can result in significant angular emission even with carefully crafted and aligned emitters. Additionally, the models used for these results were not designed to assess any coupling between these phenomena; offset extractor alignment may induce or amplify an off-axis emission site. 
    
    It also is important to remember that the area covered by the MCP is quite small compared to the plume spot size at the end of the flight tube. For our flight distance $L_{TOF}$ of 1 meter, Ø 42 mm MCP, and the half angle of roughly 15$^\circ$ observed from Fig. \ref{heatmaps}, this corresponds to approximately $1.6\%$ for our measurements. That is, every heatmap cell in the Figures in this study is generated from only 1.6$\%$ of the plume, and the MCP used in the present study is particularly large. Many other studies, e.g. \citep{schroeder2023angular,naemura2025direct,date2026electrospray}, make use of CEMs, which feature much smaller effective areas.
    

     Whole-plume compositional estimates derived from single-point sampling may then feasibly be based on random-angle, often tiny, and necessarily misrepresentative measurements. Even if the results are extrapolated to account for compositional variation within the plume based on modeling, this would perforce make the assumption that the center of the plume was correctly sampled. Extrapolations from single point measurements further assume that whatever model was used to make the extrapolation is correct, even as the dearth of angular, let alone fully two-dimensional compositional studies has made it difficult to constrain such models \citep{smith2024propagating}. Additionally, systems with only a single dimension of angular steering will be incapable of correctly measuring even the center of any given plume unless it is firing perfectly on-axis; if the beam diverges from the emitter axis at all on any axis, then it likely does on both.

    \subsection{As a Confounding Factor in Other Plume Studies}

    \citet{schroeder2023angular} performed voltage-dependent studies, finding not only the emergence of a secondary emission site at higher potential, but also that the beam wandered in the one variable axis from +7$^{\circ}$ to -8$^{\circ}$ over a 150 volt change in acceleration potential. \citet{uchizono2020emission}  similarly found a 15$^\circ$ range of firing angles between 1.2 and 1.8 kV. That the beam may wander to such a significant degree highlights the importance of obtaining full two dimensional scans to accurately recover plume content. Several studies, e.g., \citep{dandavino2011microfabrication,villegas2023impact,lyne2023simple}, have investigated voltage-dependent compositional changes. However, given the angular dependence of plume composition and such voltage-dependent wandering, studies without 2D analysis necessarily risk conflating these two distinct sources of compositional deviation; full two-dimensional plume measurement is thus required to disentangle true voltage-driven compositional variation within the plume.

    \citet{schroeder2023angular} also saw the formation of a secondary emission site but was unable to resolve it because it was off-axis in the direction that was inaccessible to the apparatus. Similarly, \citet{kingsleyTPPEmitter2025} noted a transient secondary emission site that evaporated over time. These secondary plumes may have their own internal structures unique from the main plume,  but even if they happen to match, their very existence further confounds attempts to study plumes without full 2D characterization.

    Electrospray emission angles are known to sometimes wander over time, even at constant operating parameters. For the present study, the beam was found to wander 2 degrees, from $X$ = -1$^\circ$ to +1$^\circ$ over the course of the roughly two hour experiment. Long-duration firing of electrospray emitters, either singly or in arrays as in \citet{krejci2017emission} often make use of TOF curves to estimate the $I_{SP}$ as a function of time. However, the compositional TOF data taken during such measurements would be impossible to disentangle from simple beam wandering, as the wandering would necessarily induce changes in the measured TOF curve. It also complicates transient beam studies, which by definition study unstable plumes which may wander significantly during transient voltage spikes or startup \citep{courtney2019high}. 
    
    It is also interesting to note that in the present study, the monomer-to-dimer ratio ranges from 29\% at the center to 63\% at 14$^\circ$ off-center. This ratio is observed to vary significantly across published literature, ranging from less than unity \citep{lozano2005efficiency} to significantly higher than unity \citep{Legge2011porousmetals}. In \citet{ulibarri2025direct}, the ratio was observed to rise above unity for a 3$^\circ$ span on only one side of the plume, while \citet{schroeder2023angular} observed less than unity at the scan center but $>2$ at the plume extremities. Thus, even this basic and fundamental emission metric is confounded by angular emission properties and time- or voltage-dependent angular wandering.

    \subsection{MCP Detector Response}
    Analysis of MCP detector response over wide ranges of impactor $m/q$ \citep{mathew2022characterization} shows that slower moving heavy species have smaller secondary electron emission yields. Specifically, analysis presented in the Appendix Sec. \ref{mcp} suggests that ions have a relative detection efficiency that is roughly 10$\times$ greater than the droplets observed in the present study. Thus, ion current fractions may themselves be overrepresented in the MCP output relative to the higher mass species and droplets.

\section{Conclusion}

To our knowledge, this study represents the first fully two-dimensional compositional profile of a vacuum electrospray plume. Using TOF-MS, we find significant compositional variation across the plume, with heavy species and energetic neutrals at increased abundance at the beam center based on radial scans. Nonetheless, heavy species and neutrals accelerate along paths that diverge from the primary beam by 1.5$^\circ$ and 3.4$^\circ$ respectively. Monomers exhibit a ring structure, with intensities at a maximum approximately 10$^\circ$ off the plume center axis, and the monomer-to-dimer ratio varies from 29\% to 63\%. Individual species intensities exhibit consistent azimuthal asymmetric bright spots across two different measurement runs spanning roughly two hours of operation. 

The two-dimensional TOF studies presented here have important implications for porous needle and externally-wetted ionic liquid electrospray sources. Most importantly, the observed mass species are consistent not with pure ion regime emission, but rather with the existence of a cone-jet, which is significantly less propulsively efficient. Nonetheless, single-point sampling the plume at certain angles could lead one to incorrectly conclude that the plume was operating in a highly efficient PIR mode. TOF-derived $I_{SP}$ (and thus effective propulsive $\Delta V$) at the extremities may be as much as 6 times greater than the plume centerline.

The two-dimensional system we present would also enable a careful statistical study of plume formation angles for both externally-wetted and porous emitters. Both emitter types are typically more susceptible to small manufacturing defects which may alter the plume emission angle than capillary emitters; the meniscus at relevant flow rates is sub-micron scale, and no manufacturing technique currently available guarantees surface uniformity in this size range, especially for large arrays composed of many emitters \cite{smith2026kinetic}. Thus, the anchoring of the emission-producing meniscus on externally-wetted and porous emitters may be unpredictable, with important implications for emission plume angles and the resultant compositional measurements.

The results here certainly do not disprove the existence of externally-wetted ionic liquid emitters operating in the pure ion mode, but they do conclusively show that it is possible that they can operate in the cone-jet, perhaps deceptively. It was previously commonplace to simply assume PIR operation in this current range \citep{Lyne2024JAP,Kerber2025Part1,Kerber2025Part2}, and many studies that do measure compositional content often only sample a small, effectively random patch of it. Thus, even many studies which have good reason to believe they are observing PIR emission may simply be observing a misrepresentative part of the plume. The present study thus suggests that it is beneficial to measure whole-plume species composition, or at a minimum to verify rather than assume that the beam is being sampled in the center along both the $X$ and $Y$ axes.

\section*{Acknowledgements}
The authors would also like to thank the machinists in the LASSP Professional
Machine Shop at Cornell University for making shaft adapters for the external
motor mounts with an extremely fast turnaround time that greatly aided this study.

The authors would like to thank Dr. Ben Prince for several detailed discussions about the cone-jet high mass ion and droplet structure and for sharing his previous experimental findings which further corroborate those of this study.

The authors would additionally like to thank the Cornell Center for Materials Research (CCMR) shared instrumentation facility for enabling important aspects of this study.

\section*{Funding}
This work was enabled by the Heising Simons Foundation 51 Pegasi B Fellowship
(grant number 2024-5175) and by the NASA Space Technology Graduate Research
Opportunity Fellowship (80NSSCK23K1212).

\section*{Declaration of Interests}
The authors report no conflict of interest.

\section*{Data Availability Statement}
The authors will respond to reasonable requests for the experimental data
used in this study.

\section*{Author ORCIDs}
Zach Ulibarri, \url{https://orcid.org/0000-0002-1094-9266};
Giuliana Hofheins, \url{https://orcid.org/0000-0001-5575-2240};
Sophia Gessman, \url{https://orcid.org/0009-0009-6959-1105};
Elaine Petro, \url{https://orcid.org/0000-0002-0423-4934}

\bibliography{2_ref}

\input{SupplementalArXiV}

\end{document}

%% file: SupplementalArXiV.tex
\renewcommand{\thefigure}{S\arabic{figure}}
\setcounter{figure}{0}

\section{Appendix}

    \subsection{MCP Settings}
    As with the experiments in \citet{ulibarri2025direct}, the increased SNR enabled by the goniometric aiming of the beam typically overwhelms the ARI Corp TDC-30 transimpedance amplifier at maximum 1 kV MCP amplification. Thus, for the experiments described here, the MCP was set to 900 V to eliminate signal saturation at the beam center. This reduced voltage makes it difficult to convert the measured MCP voltage signal to received current, as Hamamatsu was unable to provide gain characteristics at anything but the nominal operating voltage. As such, MCP measurements in this work are reported as raw voltages rather than calculated currents.

    \subsection{Two Dimensional Control and TOF Setup}

    Whereas previous work required rotation of the feedthrough controls on the goniometer by hand and manual data logging, the present work features a fully-automated data scanning process. StepperOnline 23HS30-2804S stepper motors are attached via couplers to the rotational feedthrough knobs and custom machined mounting hardware. Pololu Tic 36v4 USB stepper motor drivers are powered by a 24 V 2.1 amp Mean Well LRS-100-24 supply and are controlled via a custom Python software suite. The software enables manual movement of the goniometer as well as automated scans across pre-defined angular space in fixed angular step sizes. 

    The 2 degree scan setting and two samples per cell used in the experiment provide a useful balance between run time and resolution. Higher resolutions and higher spectra counts dramatically increase the time required to perform a scan, which increases the probability of the plume wandering during the experiment.

    \subsection{Data Acquisition}
    
    The scan starts at the lowest selected negative angles in both $X$ and $Y$, and then scans in fixed steps across $X$ to the highest selected angle. At each angular cell, the software stops the goniometer, records the time of arrival to this location, and then acquires a user-defined number of oscilloscope triggers. After the last averaged spectra was recorded, the software moves the goniometer to the next cell. When the scan reaches the highest $X$ angle in the scan range, the software increments the $Y$ angle by one step and returns to the minimum $X$ angle at maximum speed without stopping, and it then scans upwards through $X$ again. While a serpentine pattern that acquires data at each $X$ on the way back down would be slightly faster, this was found to occasionally introduce small errors in the $X$ angle position through mechanical slop. The software generates an H5 file \citep{collette2013python,folk2011overview} structured by $X$ and $Y$ position in angular space. Each position contains the oscilloscope acquisitions at that location and the bounding timestamps for each acquisition.

    Separately, Kickstart was used to control a Keithley 2657A source meter unit (SMU) driving the electrospray and record the voltage and output current as a function of time. Because Kickstart records time in seconds since program start rather than in absolute time, post-processing software was then used to match the acquisition timestamps to find average output current and voltage for each acquisition, and these values were then inserted into the H5 file as an average value and an error calculated from the standard error of means (SEM).

    Custom software was used for data processing, with the actual analysis written by the authors. Claude AI was used to write much of the connecting and triggering software for the oscilloscopes, but the mechanics of the motors control software were written by the authors. Claude AI was also used to make wrapper codes for the plots in the manuscript (mostly for formatting, etc.), but again, the actual data analysis was written by the authors.

    \subsection{Laser Profilometry}
    
    A Keyence VK-X260 laser profilometer with a 10x objective and a resolution of 659 nm per pixel at the Cornell Center for Materials Research was used to image the emitter down the needle axis immediately after data was taken. This scan provided highly resolved three-dimensional structure of the precise emitter configuration, most notably the position of the emitter within the circular extractor aperture. The needle tip was offset 50 $\mu$m both laterally (left) and 70 $\mu$m vertically (above) from the center of the 2.5~mm diameter extractor aperture, and recessed 600~$\mu$m below the top face of the 1.5~mm thick extractor plate. Supplemental Fig. \ref{fig: run910profilometry} shows source profiles from the laser profilometer, including the $X$ and $Y$ profile scans across the needle's center. The needle used here has a radius of curvature of $\sim$9.6$\mu m$. Scanning electron microscope images of the needle tip are shown in Supplemental Fig. \ref{fig: B4SEM}.

    \subsection{Molecular Heatmap Measurements} \label{supp_mol}

    As in \citet{ulibarri2025direct}, current for monomers, dimers, and trimers are measured as the voltage rise between bounds identified by exponentially modified Gaussians (EMGs) fitted to the derivative of the data smoothed by a Gaussian blur (in this case of width 3 in the index domain), as shown in Fig. 2 of \citet{ulibarri2025direct}. Whereas before the found indices were averaged across the entire dataset to find a single set to use, here they were chosen from a single exemplar spectrum for expediency, as the slight changes between individual spectra do not have a meaningful impact on the results presented here. As before, the voltage rise in between each species is taken to be fragmenting species. For example, between the upper monomer bound and the lower dimer bound, the voltage rise is taken as the fragmenting dimer population. However, it should be noted that this is very much a simplification, as much higher mass cluster groups and even droplets may contribute ions in these regions as a result of collisions \citep{Smith_2025}. As before, the voltage rise from the upper trimer bound to an arbitrary point called the high mass cutoff at the end of the recorded data is taken as the heavy particle or droplet yield. The high mass cutoff is now 13,000 AMU (as opposed to 2800 AMU in the previous study) for the two dimensional TOF and $5\times 10^7$ amu/q in the single TOF at plume centerline study.

    Whereas in our previous study the MCP voltage baseline was treated as a confounding factor and eliminated through subtraction, here we recognize it as an important fundamental feature of the emission. The MCP baseline is shown to have significant two-dimensional variation, roughly centered on the MCP signal's intensity, suggesting that significant populations of high-velocity neutral particles are present within a narrowly-defined region of the plume. In principle, the feature could also be the result of secondaries or any other particles that are not deflected by the high voltage TOF gate. To calculate the MCP baseline, new set of bounds called the baseline bounds were arbitrarily placed before the monomer lower bound to determine the average MCP bias for each spectrum. The bounds are chosen very near each other close to the monomer signal because in the high-off-axis spectra, a small voltage drop is observed near the gate pulse at the beginning of the data. The placement of these bounds was visually selected to ignore this confounding factor and produce a uniform baseline for all of the data.

    It should also be noted that a ``true'' MCP bias of 3.2 mV was observed at high-off-axis angles, which is thought to be an electronics artifact or the result of secondary species impacting chamber walls. This value was subtracted from the TOF curve before any other processing was done to produce data centered at the ``true'' zero voltage and provide a more accurate measurement of the neutral plume contribution. The subtraction does not meaningfully alter any of the results presented here, except for quantification of the energetic neutral population and monomer yield, both of which require this subtraction for accurate results.

     \subsubsection{Error Bars}\label{error}

    There are two sources of statistical variation which contribute to the error bars given in the species yield plots: azimuthal variation and multiple measurements. Because 72 slices were taken in each of the two Runs, the $\frac{1}{\sqrt{n}}$ factor produces artificially small error bars that obfuscate the true scale of the statistical variation. This is because the small separation of only $5^\circ$ between each ray on the smoothed data sometimes produces significant correlation between adjacent samples, meaning that the true or effective sample count $n$ is lower than 72.

    To correct for this, we estimate an effective sample size $n_\mathrm{eff}$ from the
    lag-1 angular autocorrelation $\rho$ of the intensity values around each sampled ring at radius $r$:
    \begin{equation}
    n_\mathrm{eff}(r) = n \cdot \frac{1 - \rho(r)}{1 + \rho(r)}
    \end{equation}
    \noindent where $\rho(r)$ is the Pearson correlation coefficient between intensity
    values at adjacent azimuthal positions. This is a standard error estimation method for auto-correlated measurements \citep{bayley1946effective}, and it has the effect of reducing $n_\mathrm{eff}$ when neighboring rays measure nearly identical intensities, but approaches $n$ when they are highly differentiated. The azimuthal SEM is then computed as $\sigma_\mathrm{azimuthal}(r) = \sigma(r) / \sqrt{n_\mathrm{eff}(r)}$, where $\sigma(r)$ is the standard deviation of intensities across all 72 rays at radius $r$.

Results from individual runs are reported only in the Supplemental, with error
bars generated by this corrected azimuthal SEM. The error bars in
Fig.~\ref{radial} combine azimuthal and measurement variation, with the SEM of
both sources added in quadrature as
\begin{equation}
    \sigma_\mathrm{total}(r) = \sqrt{\sigma_\mathrm{azimuthal}^2(r) +
    \sigma_\mathrm{measurement}^2(r)}
\end{equation}
\noindent to ensure that both sources of variation are accounted for.

\subsection{MCP Detector Response to varying $m/q$}\label{mcp}
The microchannel plate (MCP) detector employed for ion arrival time measurements operates on the principle that energetic species incident upon microchannel tubes, coated with a material of known secondary electron emission (SEE) yield, initiate a collision cascade within the channel. This cascade results in the sequential multiplication of secondary electrons, producing an amplified charge signal that is subsequently measured via oscilloscope. However, it is known that MCP response is dependent on impacting particle mass, velocity, and charge state \citep{mathew2022characterization, liu2014detection, gilmore2000ion}. Namely, heavier and thus slower particles are known to have lower secondary electron yields and detection efficiencies \citep{gilmore2000ion}.

In this ionic liquid electrospray plume, there exists 7 decades of $m/q$ ratio species. Thus, the varying detection efficiencies must be taken into account when delineating species plume fractions. While the droplet populations have much lower velocities (minimum 300 m/s) compared to ions (30 km/s), the comparative detection efficiency should be low. However, work by \citet{mathew2022characterization} shows MCP response not only dependent on mass and velocity, but takes into account the charge state as well. This work derives a secondary electron yield ($\gamma$), which is directly proportional to the output current recorded on the oscilloscope, as
\begin{equation}
    \gamma \propto m^{0.28} v^{1.54} \propto m^{-0.49} z^{0.77}
\end{equation}
where $m$ is the species mass, $v$ is the impact velocity, and $z$ is the number of unit elementary charges the impacting species possesses. The exponents were empirically derived from data fits \citep{mathew2022characterization}. This shows that when a species possesses a large number of charges, this can offset the loss in yield due to a large particle at a slow speed. Assuming the ILIS employed in these tests is operating in cone-jet emission, \citet{miller2021capillary} estimates that the `large $m/q$' population of an EMI-Im capillary at a 0.72 nl/s flow rate consists of droplets ${\sim}1.7 \times 10^{8}$ AMU with ${\sim}500$ charges. Compared to the `low $m/q$' population, where \citet{miller2021capillary} reports $m \sim 1.4 \times 10^{7}$ AMU and ${\sim}100$ charges, the relative detection efficiency is essentially equal. The lighter and faster progeny droplets in the lower $m/q$ range possess a roughly equivalent detection efficiency due to their proportionally reduced charge relative to the heavier, slower large $m/q$ primary droplets.

The surprising result is that ions have a relative detection efficiency approximately 
10$\times$ that of the droplet populations, suggesting that ion current fractions may 
be significantly overrepresented in the MCP output relative to the true plume composition. This does not imply a simple factor of 10 correction to the ion fraction, as the true 
ion current fraction depends on the absolute values of the secondary electron yield 
for each species, which are functions of additional instrument-specific parameters and 
operating conditions not characterized here. Within the droplet population itself, 
however, the detection efficiency is remarkably consistent across the full droplet 
$m/q$ range --- the charge state scaling with mass from jet breakup naturally 
compensates the mass penalty, such that relative droplet population fractions are 
well represented in the measured signal. The species most susceptible to undercounting 
are those with anomalously high mass and low charge state relative to the nominal 
charge-mass scaling of jet breakup.
    
\section{Supplemental Images}
  \begin{figure*}[h!]
    \centering
    \includegraphics[width=\textwidth, scale=1]{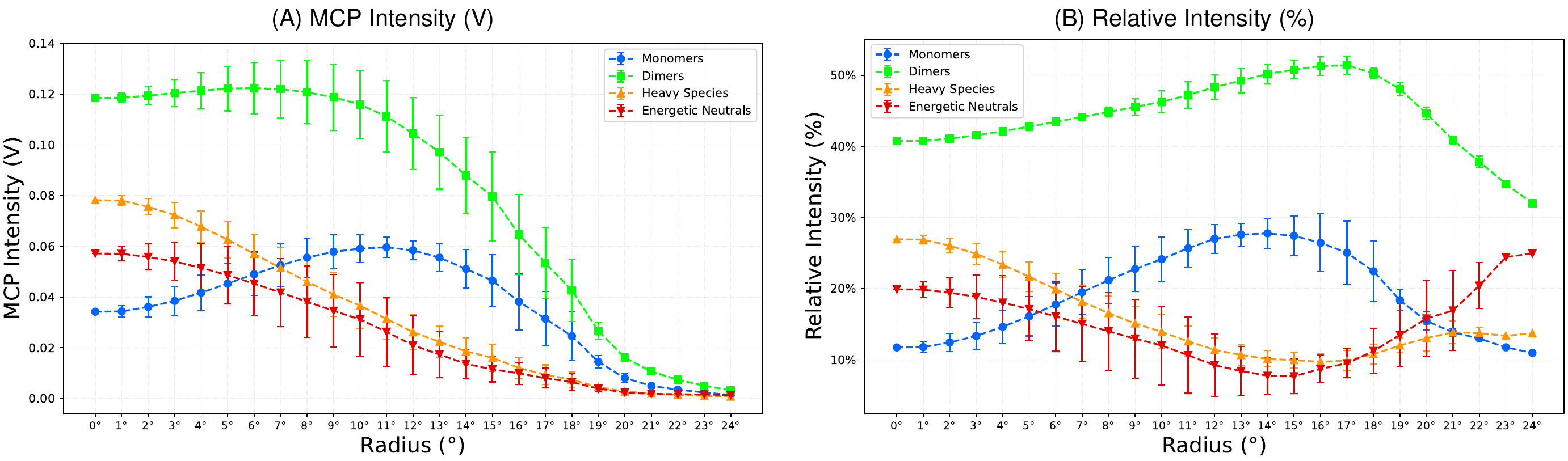}
    \caption{The yields of each molecular species as a function of radius from the plume center (not the needle axis) for Run A. Error bars represent solely the azimuthal variation of the plume as described in Sec. \ref{error}.}
    \label{Run2A} 
    \end{figure*}

    \begin{figure*}[h!]
    \centering
    \includegraphics[width=\textwidth, scale=1]{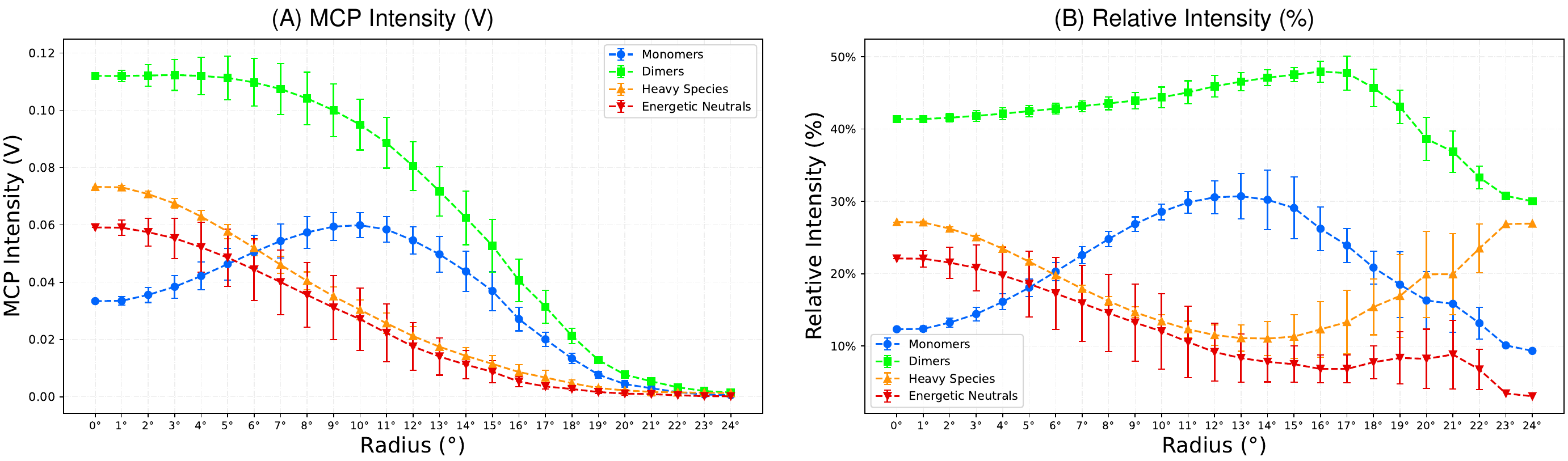}
    \caption{The yields of each molecular species as a function of radius from the plume center (not the needle axis) for Run B. Error bars represent solely the azimuthal variation of the plume as described in Sec. \ref{error}.}
    \label{Run2B} 
    \end{figure*}

\begin{figure*}[h!]
\centering
\includegraphics[width=\textwidth, scale=1]{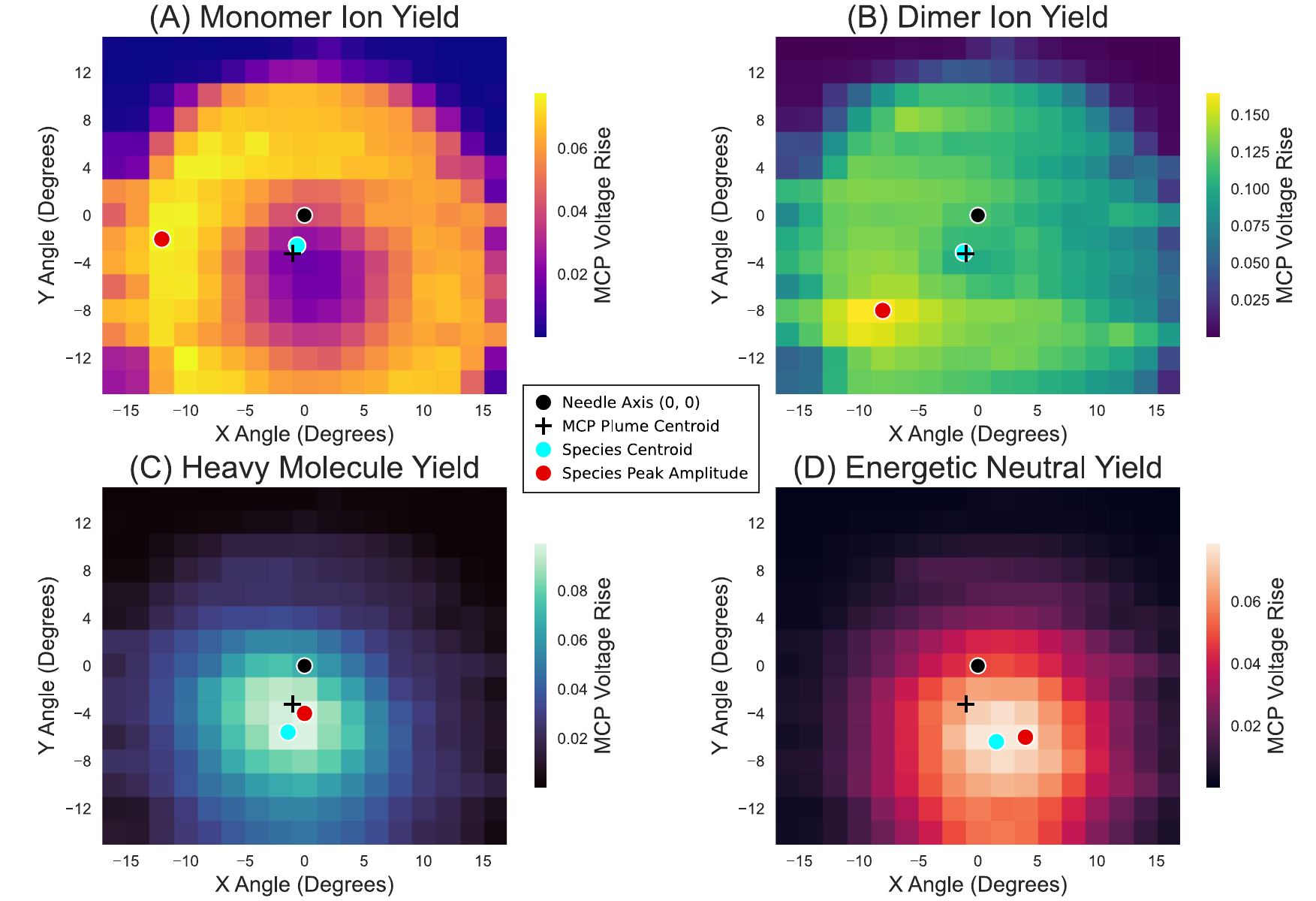}
\caption{Heatmaps of the species yields for Run A. This plot is identical to the in-text Fig. \ref{heatmaps}, but is repeated here for rapid comparison to Fig. \ref{R2B}.}
\label{R2A}
\end{figure*}

\begin{figure*}[h!]
\centering
\includegraphics[width=\textwidth, scale=1]{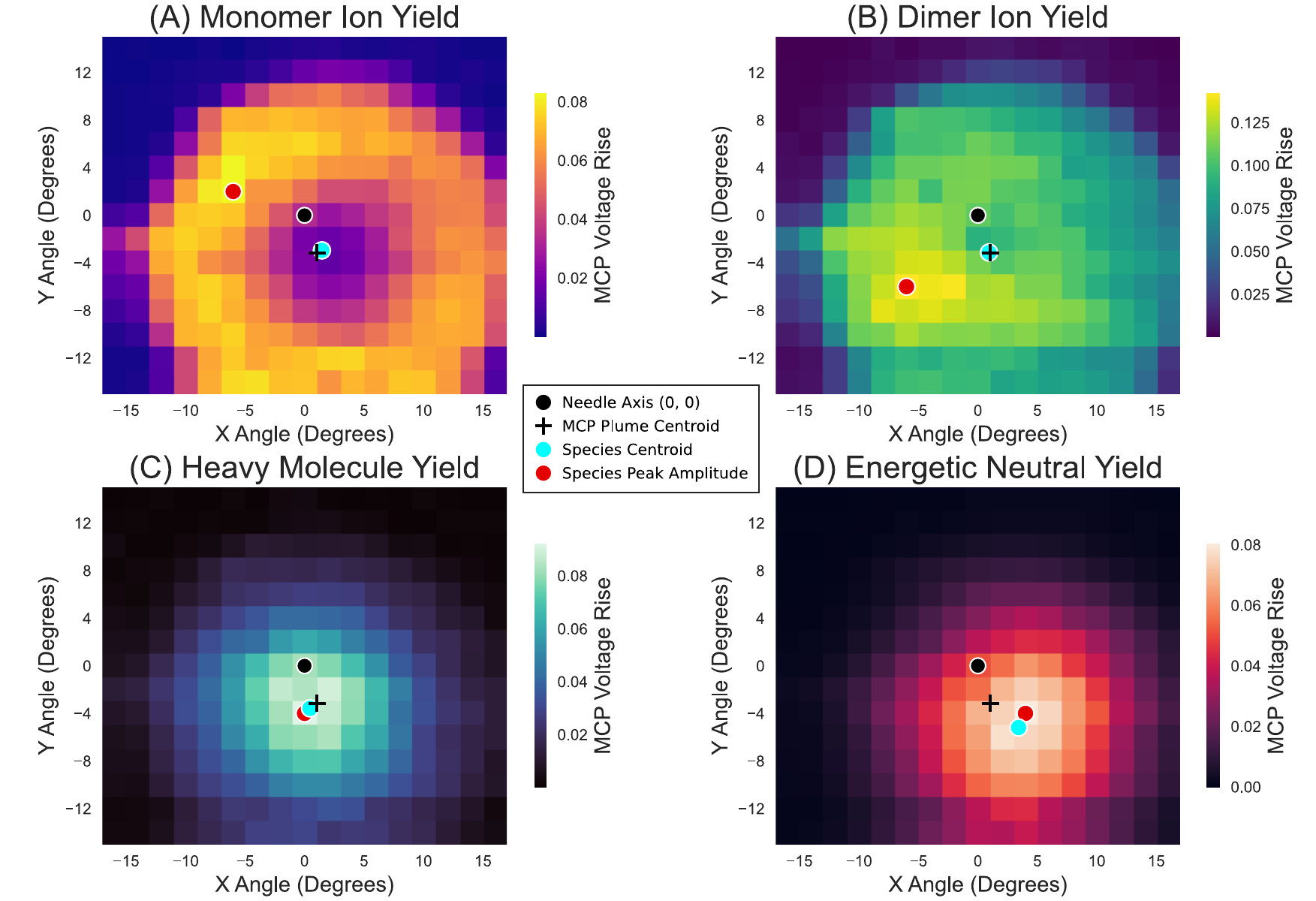}
\caption{Heatmaps of the species yields for Run B.}
\label{R2B}
\end{figure*}

\begin{figure*}[h!]
\centering
\includegraphics[width=\textwidth, scale=1]{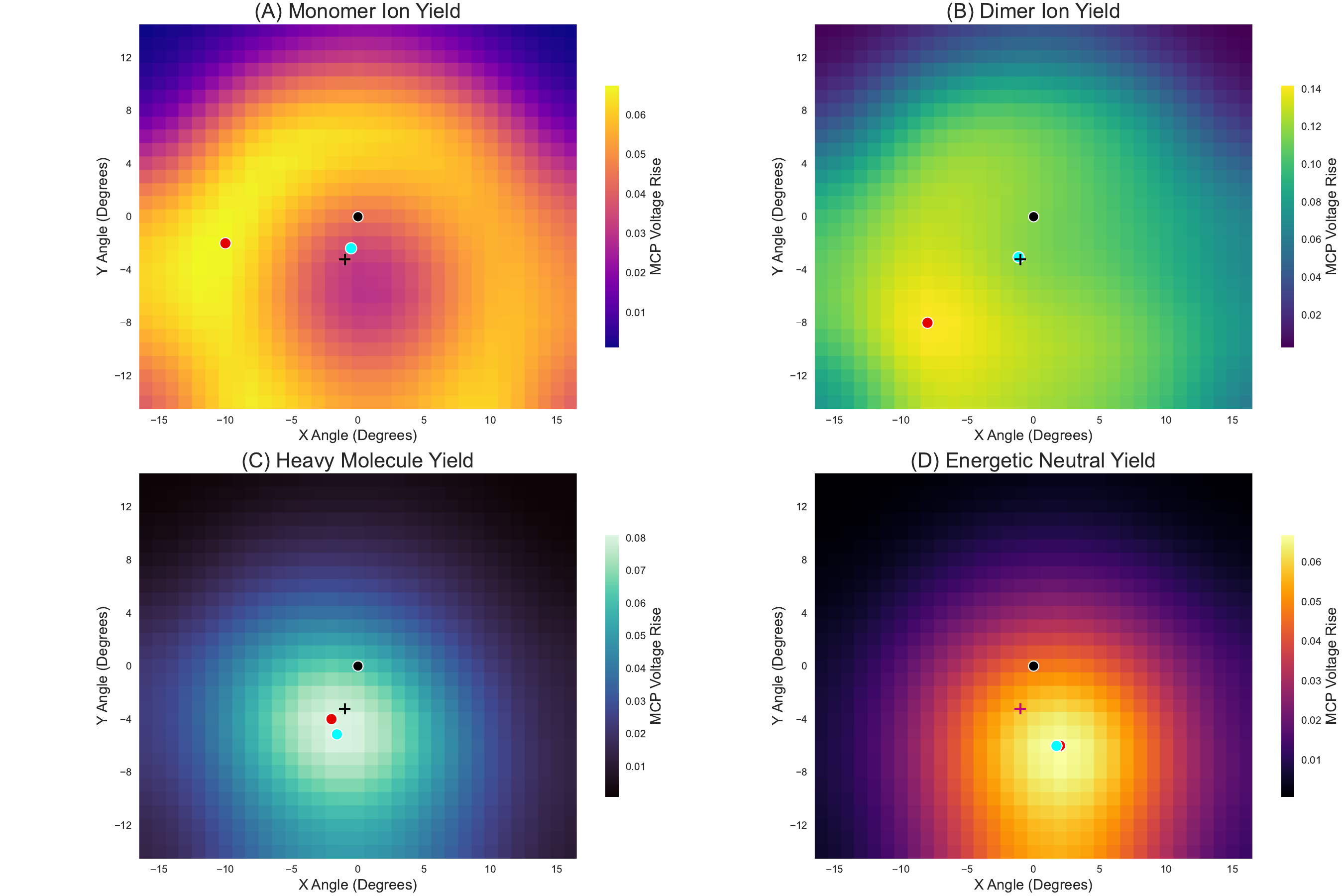}
\caption{Heatmaps of the species yields for Run A, smoothed by a 2D Gaussian filter as described in Sec. \ref{statistics}.}
\label{R2A-smoothed}
\end{figure*}
\begin{figure*}[h!]
\centering
\includegraphics[width=\textwidth, scale=1]{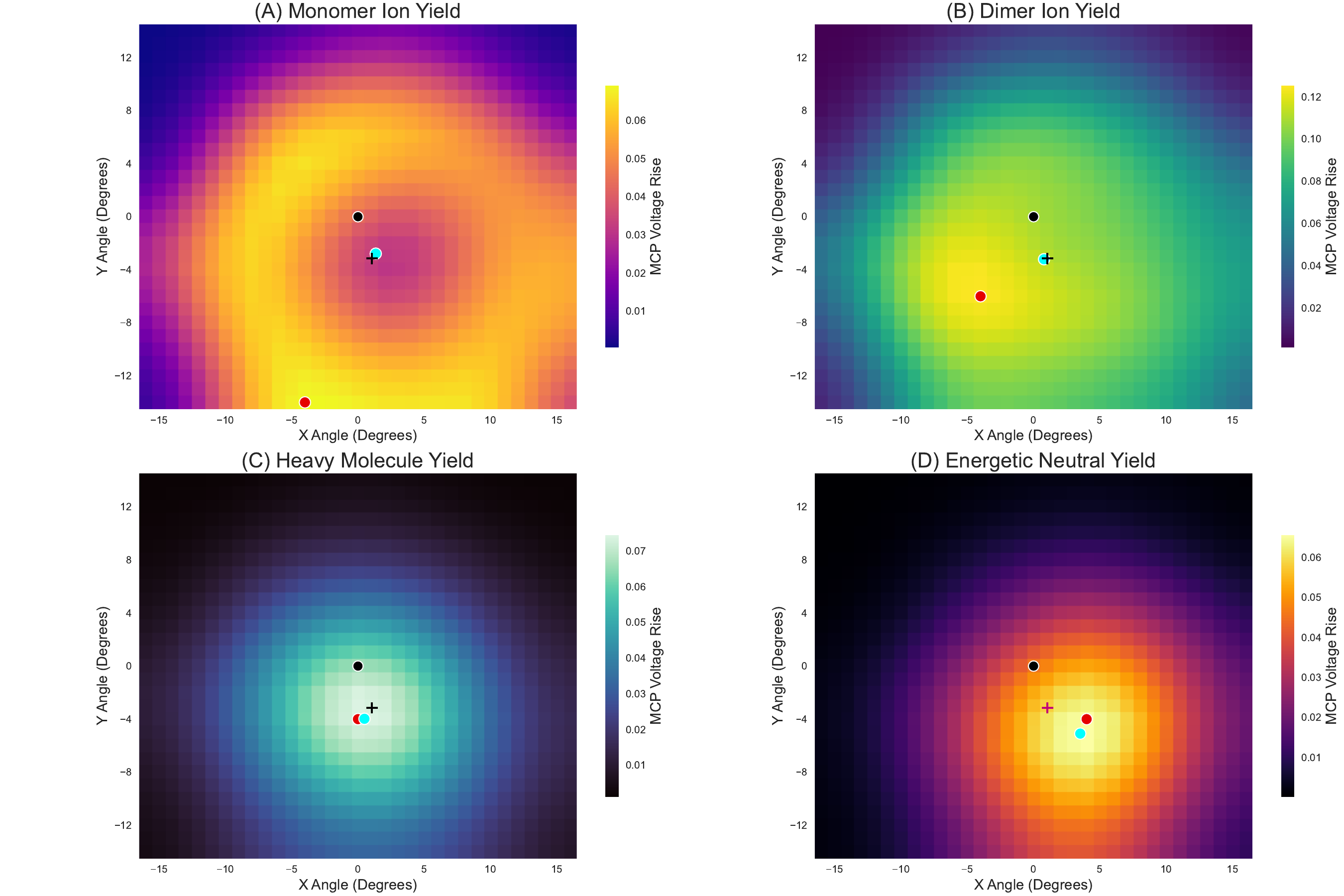}
\caption{Heatmaps of the species yields for Run B, smoothed by a 2D Gaussian filter as described in Sec. \ref{statistics}.}
\label{R2B-smoothed}
\end{figure*}

\begin{figure*}[h!]
\centering
\includegraphics[width=\textwidth, scale=1]{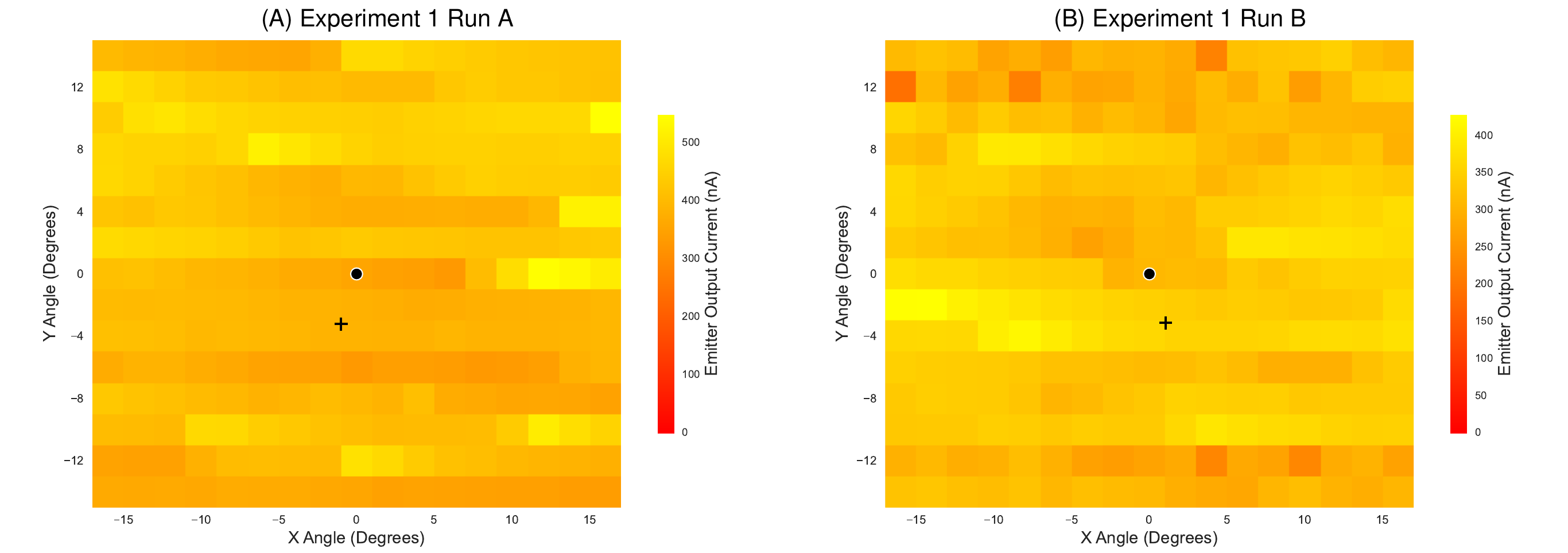}
\caption{The source meter unit (SMU) output current for Runs A and B. It can be seen that the current was largely stable during the experiments, and an increase in emission current cannot explain the asymmetries observed in the species yield data.}
\label{output}
\end{figure*}

\begin{figure*}[h!]
\centering
\begin{minipage}{0.7\textwidth}
    \centering
    \includegraphics[width=\textwidth]{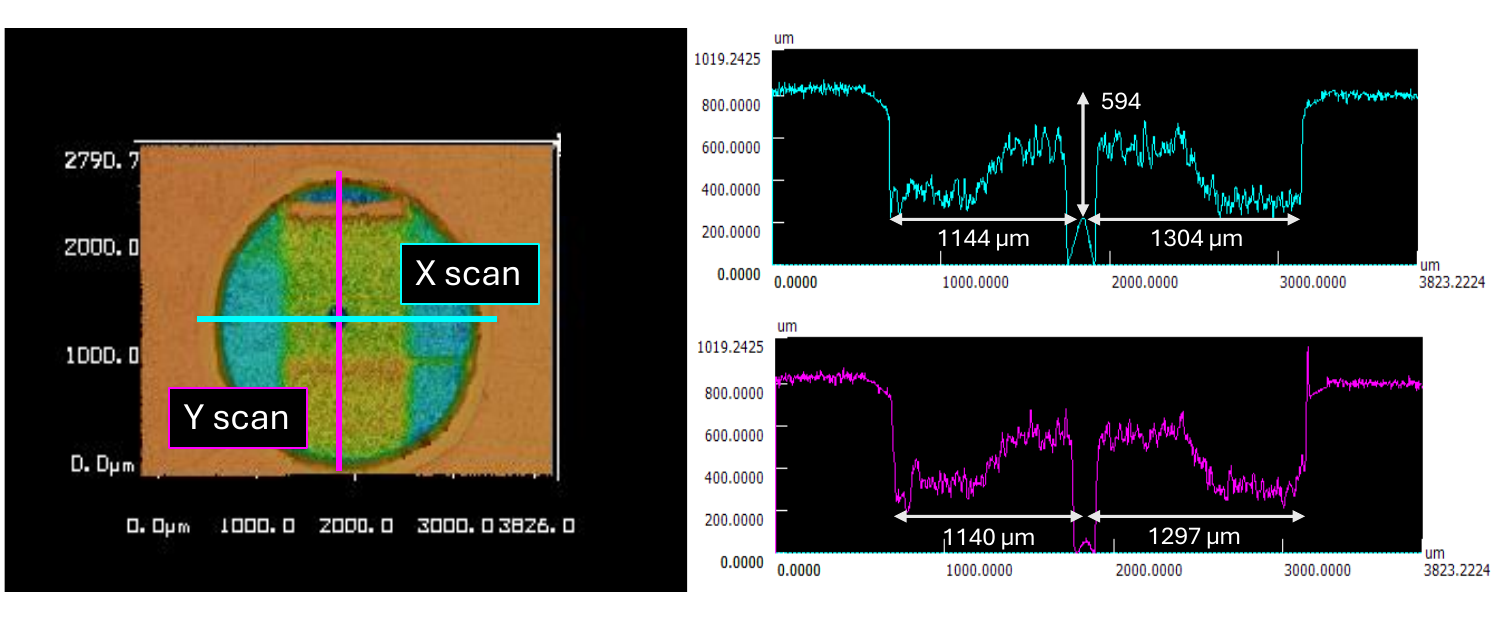}
\end{minipage}%
\hfill
\begin{minipage}{0.3\textwidth}
    \centering
    \includegraphics[width=\textwidth]{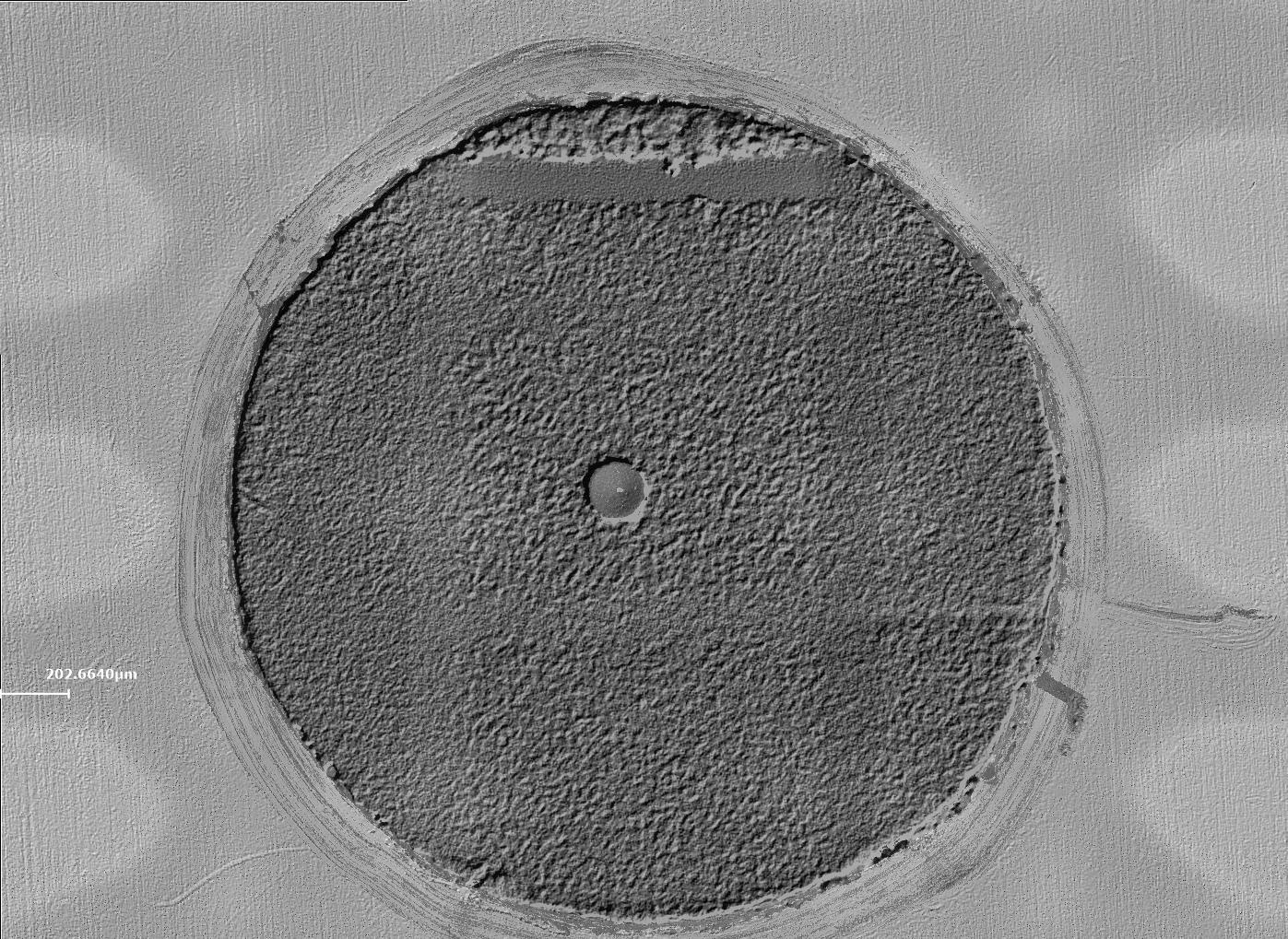}
\end{minipage}
\caption{Profilometry of ion source centering in extractor aperture.}
\label{fig: run910profilometry}
\end{figure*} 

\begin{figure*}[h!]
\centering
\includegraphics[width=\textwidth, scale=1]{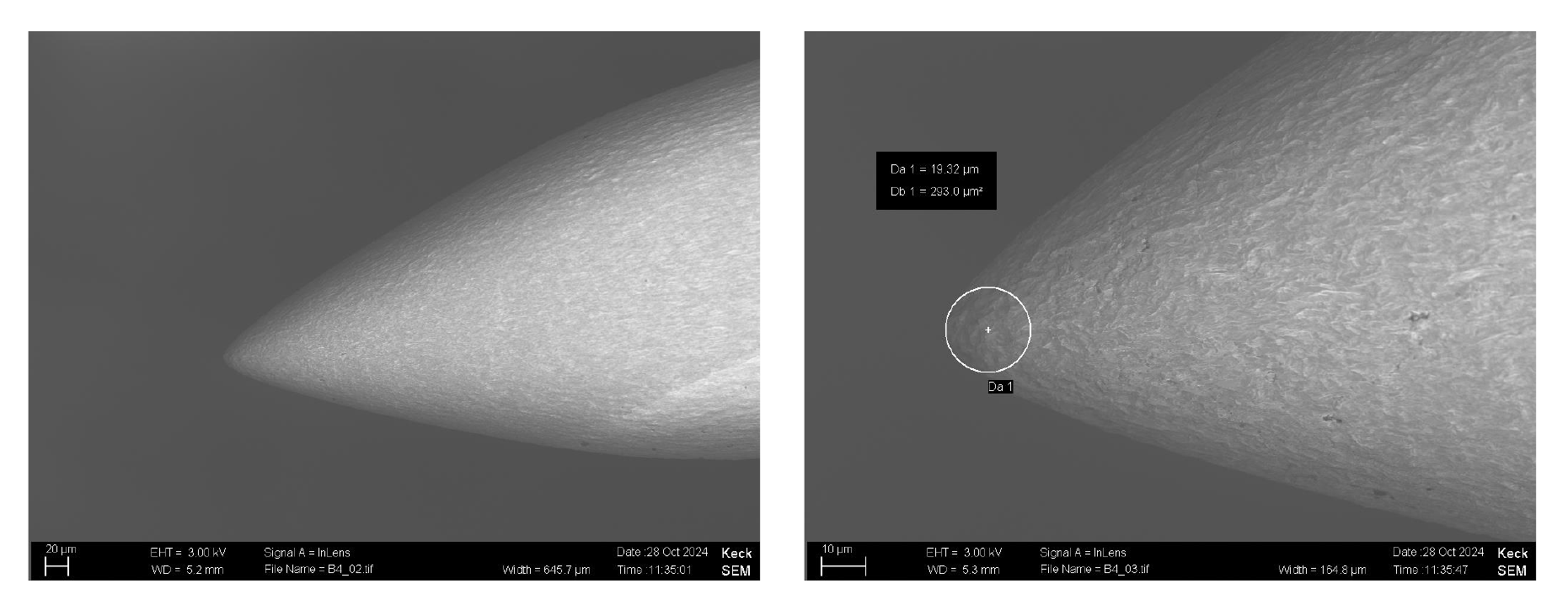}
\caption{Scanning electron microscope images of the externally-wetted tungsten ion source. }
\label{fig: B4SEM}
\end{figure*}